\documentclass[11pt]{article}
\usepackage{graphicx}

\newcommand{\BABARPubYear}    {01}

\newcommand{\BABARProcNumber} {64}
\newcommand{\SLACPubNumber} {9077}

\input{babarsym.tex}

\long\def\inst#1{\par\nobreak\kern 4pt\nobreak
    {\it #1}\par\vskip 10pt plus 3pt minus 3pt}

\begin{document}
{\pagestyle{empty}

\begin{flushright}
SLAC-PUB-\SLACPubNumber \\
\babar-PROC-\BABARPubYear/\BABARProcNumber \\
%\babar-PUB-\BABARPubYear/\BABARPubNumber \\
%hep-ex/\LANLNumber \\
December, 2001 \\
\end{flushright}

\par\vskip 4cm

% Title of the paper
\begin{center}
\Large \bf  NEW RESULTS FROM \babar
\end{center}
\bigskip

\begin{center}
\large 
Aaron Roodman\\
Stanford Linear Accelerator Center \\
Stanford, California, 94309.  \\
(for the \lbabar\ Collaboration)
\end{center}
\bigskip \bigskip

% Abstract
\begin{center}
\large \bf Abstract
\end{center}
  The \babar experiment at the PEP-II asymmetric B factory at SLAC
  has collected a large sample of data at the $\Upsilon(4S)$
  resonance. I will summarize \babar's new results on CP violation,
  \B mixing and lifetimes, and a selection of rare \B decays.  In
  particular, I will describe in detail the measurement of the CP
  violating parameter \stwob; \babar has observed CP violation
  in the neutral \B system finding
  $\stwob = 0.59 \pm 0.14 \pm 0.05$.  

\vfill
\begin{center}
Contributed to the Proceedings of the 21$^{th}$ Physics in Collision Conference, \\
6/28/2001---6/30/2001, Seoul, Korea
\end{center}

\vspace{1.0cm}
\begin{center}
{\em Stanford Linear Accelerator Center, Stanford University, 
Stanford, CA 94309} \\ \vspace{0.1cm}\hrule\vspace{0.1cm}
Work supported in part by Department of Energy contract DE-AC03-76SF00515.
\end{center}

\section{Introduction}

There are four known manifestations of matter versus anti-matter
asymmetry. The first is the observed lack of anti-matter in the
universe.  The second and third are the presence of CP violation in
decays of the neutral Kaon, in mixing and in decays. The fourth is
observation of CP violation in decays of the neutral $B$ meson, the
primary subject of this paper.  The fundamental goal of the \babar
experiment is to understand the relationship between these four
observations.  

CP violation is one of the conditions which must be present in the
early universe to create the cosmological baryon asymmetry.  One of
the great ironies in particle physics is that the observed CP
violation in the CKM matrix of the Standard Model of the
charged-current weak interaction is too weak by many orders of
magnitude to explain the baryon asymmetry.  Thus the goal at
\babar is to discover whether the CKM matrix is the source of all
CP violation in the $B$ meson.  If there is a component of the observed
CP violation which is not explained by the CKM matrix, then perhaps
this extra component may play a role in the baryon asymmetry.  In
addition, there is presently no deeper understanding of the
fundamental source of the CKM matrix, with its striking hierarchy of
couplings, and a first step towards this is to thoroughly measure the
size and phases of the quark's weak couplings.

This paper will describe some of the recent results from the \babar
experiment. For results from the similar experiment in Japan, BELLE,
operating at the KEK-B accelerator, please see the relevant paper in
these proceedings. 

\subsection{CP Violation and the CKM Matrix}

In the Standard Model the weak charged couplings between quarks are
given by a 3x3 unitary matrix, the CKM Matrix.  In this model all CP
violation is due to a single complex phase in the CKM matrix. In the convenient
parametrization due to Wolfenstein\cite{wolfen}, the phase is placed in the
$V_{td}$ and $V_{ub}$ elements; 
%\begin{equation}
%  V_{ij} =        \begin{pmatrix}
%        1-\lambda^{2}/2 & \lambda & A\lambda^{3} (\rho-i\eta) \\[0.15\semcm]
%        -\lambda & 1-\lambda^{2}/2 & A\lambda^{2} \\[0.15\semcm]
%        A\lambda^{3}(1-\rho-i\eta) & -A\lambda^{2} & 1 \\
%       \end{pmatrix}
%\end{equation}
the phase of $V_{td}$ is $\beta$, and the phase of
$V_{ub}^{*}$ is $\gamma$.  Interference between \Bz mixing and decay
allows us to measure \stwob.  A value of \stwob around 0.7 is expected
from the constraints of  measurements of CP violation in the
neutral Kaon,  $B_{d}$ and $B_{s}$ mixing, and the ratio
$|V_{ub}/V_{cb}|$ using $b \to u$ and $b \to c$ semi-leptonic decays.
Unfortunately, these measurements suffer from significant theoretical
uncertainty.  The ultimate goal is a precise measurement of several
different manifestations of CP violation in the \B system.

% The constraints due to the unitarity of this
%matrix are commonly expressed in the $\rho-\eta$ plane, shown in
%Figure~\ref{fig:triangle}.  Outside of our measurement of CP
%violation, there are four experimental constraints on th,
%also shown in Figure~\ref{fig:triangle}.  Measurements of the ratio
%$|V_{ub}/V_{cb}|$ using $b\rightarrow u$ semi-leptonic decays
%constrain $|\rho - i \rho|$, $B_{d}$ and $B_{s}$ mixing measurements,
%discussed in detail below, constrain $|1-\rho-i\eta|$, and the
%measurement of CP violation in the neutral kaon constrains the
%combination $\eta\left[ (1-\rho) + 0.33 \right] = 0.40$.
% rosner has this ias 0.35 instead of 0.33 and 0.48\pm 0.20 instead of
% 0.40

\subsection{PEP-II Asymmetric B Factory}

The \babar experiment operates at the PEP-II asymmetric \B factory at
SLAC.  This accelerator consists of two separate storage rings, with
positrons at an energy of 3.1~GeV and electrons at 9.0~GeV, using
the SLAC linac as an injector. PEP-II operates at the $\Upsilon(4S)$
resonance, with the center-of-mass 
boosted by  $\beta\gamma = 0.55$ enabling \babar to measure time
dependent asymmetries.
PEP-II has now surpassed its design
luminosity goal of $3\times 10^{33} {\rm cm}^{-2}{\rm sec}^{-1}$, with
typical peak luminosities of $4.2\times 10^{33}$.  Recently the  integrated
luminosity has surpassed $240 \invfb$ per day.  Design and typical
parameters for PEP-II are shown in Table~\ref{table:pepii}.  The
integrated luminosity over the course of the last two years is shown 
in Figure~\ref{fig:luminosity}. All results, except for \stwob, are
based on a data sample of roughly $23 \times 10^{6}$ \BB decays.  The \stwob
result is based on a sample of $33 \times 10^{6}$ \BB decays.

\begin{table}[thb]
  \centering
  \caption{ \it Design parameters and typical values for PEP-II. }
  \vskip 0.1 in
  \begin{tabular}{|lcc|} \hline
    Parameter & Design & Typical \\
    \hline
    \hline
    $I_e^{+}$ (mA) &  2140 & 1590 \\  
    $I_e^{-}$ (mA) &  750 & 950\\  
    $N_{Bunches}$ & 1658 & 728 \\
    Tune shift & 0.03 & 0.07 \\
    Vertical Spot ($\mu m$) & 5.4 & 5-6\\ 
    Peak $\mathcal{L}$ ($10^{33}$) & 3.0 & 4.2 \\
    Daily $\int \mathcal{L}$ ($\invpb$) & 135 & 240 \\ 
    \hline
  \end{tabular}
  \label{table:pepii}
\end{table}

\begin{figure}[thb]
  \begin{center}
  \includegraphics[bb=60 130 550
  660,clip=true,height=0.5\textheight]{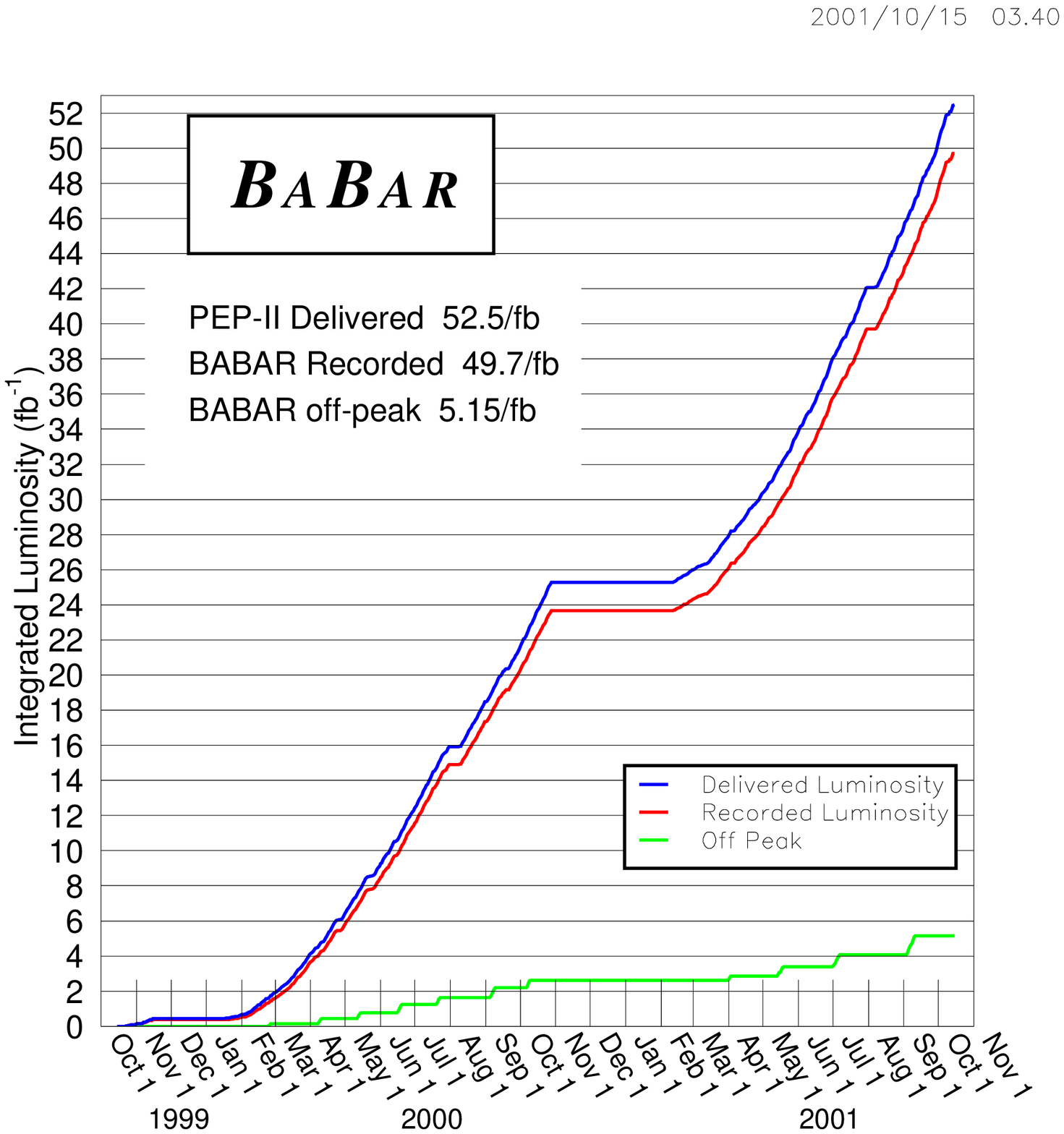}
  \end{center}
  \caption{\it
    Integrated Luminosity as a function of time.
    \label{fig:luminosity} }
\end{figure}

There are several important factors which allow PEP-II to achieve high
luminosity.  First, separate electron and positron rings allow high
currents to be stored without disturbing the {\it other} beam.
Second, having separate rings allows for the storage of many bunches,
which permits high currents without large beam-beam tune shifts.  In
the simplest terms, the beam-beam tune shift describes the amount that
a beam is perturbed by another beam. This tune shift is due to Coulomb
interactions and is proportional to the number of particles in the
other beam.  An accelerator cannot function stably with too large a
value of this tune shift\cite{lee}.  Next, the use of numerous
feedbacks makes stable operation at high luminosity possible.  For
example, PEP-II has single bunch feedbacks on the longitudinal and
transverse position of every bunch, as well as slow feedbacks, of
roughly 1-2 Hz, on the beam orbit, interaction angle, and luminosity
itself.\cite{Hendrickson:2000ma,himel}  In practice the luminosity at
PEP-II is limited by the amount of RF power available in each ring,
and by beam-component heating, mostly from higher order RF modes.

\subsection{\babar Detector}

The \babar detector\cite{babar} is shown in 
Figure~\ref{fig:babar}.  The detector consists of a five layer
double-sided silicon vertex detector (SVT), a 40 layer drift chamber
(DCH), a cherenkov detector with quartz radiators and PMT readout
(DIRC), a superconducting solenoid, a CsI(Tl) electromagnetic
calorimeter (EMC), and an iron flux return instrumented with 19 layers
of resistive plate chambers (IFR). The SVT has $15\mu m$ single hit
resolution and the impact parameter resolution is $\sigma_{z} = 65 \mu
m$ at 1~GeV.  The DCH has a momentum resolution of  $\sigma_{p_{t}} =
0.13 p_{t} + 0.45 \,\%$ and a dE/dx resolution of $8\%$ for high
momentum electrons.  The DIRC provides excellent particle
identification, with $2.5 \sigma$ K/$\pi$ separation at 4~GeV. The EMC
has an energy resolution of  $\sigma_{E}/E = 1.85  \oplus
2.32/\sqrt[4]{E} \,\%$. Finally, the IFR provides $\mu$ and $\KL$
identification. 

\begin{figure}[thb]
  \includegraphics[width=0.95\textwidth]{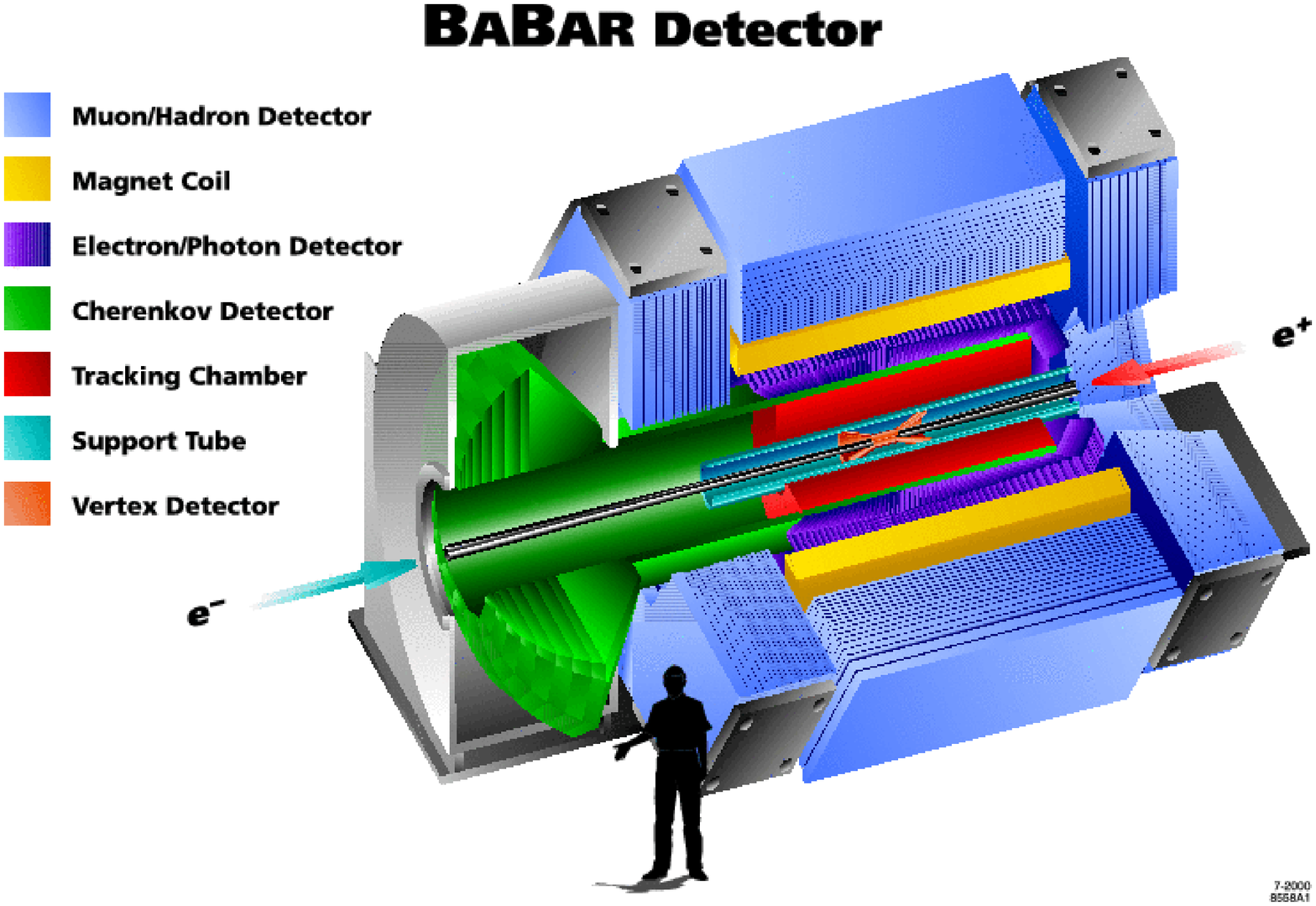}
  \caption{\it
    \babar detector.
    \label{fig:babar} }
\end{figure}

\section{CP Violation Measurement}

In this section, we will discuss in detail \babar's recent
measurement of CP violation in the neutral \B meson.

\subsection{CP Violation and Mixing in the \B meson}

Typically CP violating amplitudes are measured through the
interference of at least two amplitudes.  In our case, the amplitude
for the \Bz's decay to a final state which is also a CP eigenstate
interferes with the amplitude for the \Bz to mix into a \Bzb and
then decay into the same final state.  The CP violating phase may occur in 
either the mixing amplitude for $\Bz \leftrightarrow \Bzb$ or the
decay amplitude for $\Bz \to f_{CP}$, or both.  

The angle $\beta$ of the CKM matrix may be measured using the
final state $b \rightarrow c\bar{c}s$; the dominant decay mode of this
type is \bpsiks.  In the Wolfenstein
parametrization, this decay amplitude has no phase, as seen in the
tree-level diagram in Figure~\ref{fig:diagram}a.  In addition there is
a penguin diagram, shown in Figure~\ref{fig:diagram}b, which also has
zero phase, for the dominant contribution with a top quark in the
loop. Thus, unlike the case for other CP eigenstate final states,
there is no extra interference between tree and penguin diagrams for
$b \rightarrow c\bar{c}s$, and no appreciable theoretical uncertainty.

\begin{figure}[thb]
  \includegraphics[bb=90 540 325 700, clip=true,width=0.45\textwidth]{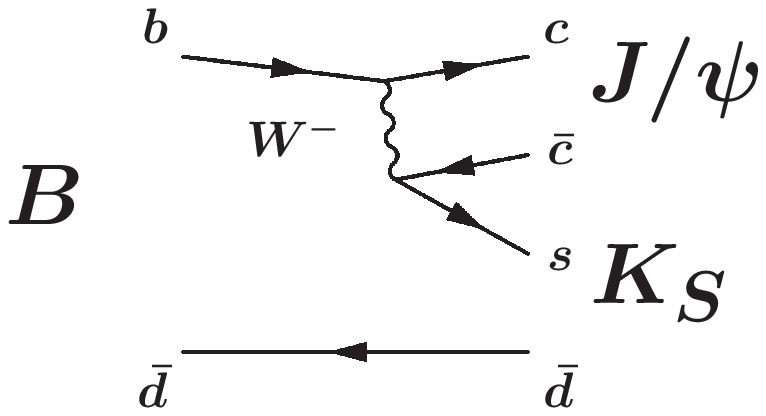}
  \includegraphics[bb=90 540 325 700, clip=true,width=0.45\textwidth]{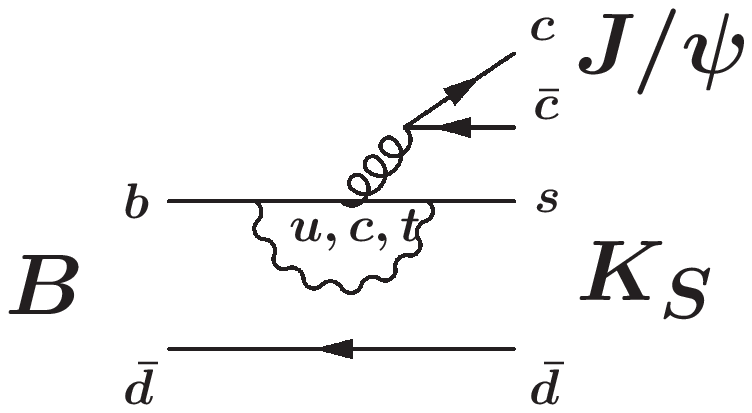}
  \caption{\it
    Feynman diagrams for \bpsiks: a) tree level
    and b) penguin.
    \label{fig:diagram} }
\end{figure} 

At the $\Upsilon(4S)$ resonance we make \Bz mesons in correlated
pairs, and both \B mixing and CP violating asymmetries depend on the
decay time difference \deltat, not the absolute decay times.  If both
decay at the same time, then one must decay as a \Bz and the other as
a \Bzb.  However, if the decay occurs at different times, then \BzBzb
mixing may yield two \Bz or two \Bzb decays. Thus the CP violating
asymmetry from the interference between mixing and decay amplitudes
will increase as the mixing probability increases, and there will be
no asymmetry at $\deltat = 0$.

The time-dependent CP violation measurements are made with
one decay to a {\bf CP} eigenstate and the other to a {\bf flavor}
eigenstate.  The decay rate is given by
\begin{equation}
      f_{\pm}(\deltat) =  \frac{e^{-|\deltat|/\tau_{\Bz}} } 
                          {4 \tau_{\Bz} } \times        \left[ 1  \pm 
           (1-2\omega) \mathcal{I} m \lambda_{f_{CP}} \sin (\deltamd
\deltat) \right] \otimes \mathcal{R}(\deltat)  
\end{equation}
for \Bz and \Bzb flavor eigenstates. In the standard model  $\lambda_{f_{CP}} =
\pm e^{i2\beta}$ for $b \rightarrow c\bar{c}s$ decays. Here
$\omega$, the mistag rate, is  the
probability of incorrectly identifying the flavor eigenstate, and
$\mathcal{R}$ is the
\deltat resolution function.
%\begin{equation}
% \lambda_{f_{CP}} = \eta_{f_{CP}} \frac{p}{q}
% \frac{\bar{A}_{\bar{f}_{CP}}}{A_{f_{CP}}} 
%\end{equation}
%and $\eta_{f_{CP}}$ is the intrinsic CP of the final state.  
The \Bz mixing measurements are made with both \Bz decaying into
flavor eigenstates.  The decay rate is given by
\begin{equation}
   f_{\pm}(\deltat) = 
   \frac{\exp{-|\deltat|/\tau_{\Bz}}}{4\tau_{\Bz}}
   \times \left[ 1 \pm (1-2\omega) \cos{\deltamd \deltat} \right] 
    \otimes \mathcal{R}(\deltat)
\end{equation}
for un-mixed and mixed events, and
where $\deltamd$ is the mass difference between the CP Eigenstates
$B_{H}$ and $B_{L}$. Since the mistag rate and the   \deltat
resolution function can be measured directly using the mixing sample, the CP
asymmetry measurement is immune from most systematic uncertainties.

\subsection{Elements of the CP Violation Measurement}

To measure \stwob we identify a sample of decays of the sort
\bpsiks, determine the flavor of the other \Bz
through its decay into a flavor eigenstate, and measure the time
difference \deltat between the two decays.  In the center-of-mass
typically $\deltaz \sim 30 \mu m$, but the asymmetric collider boosts
this to  $\deltaz \sim 260 \mu m$ in the lab frame, which is roughly twice
the vertex resolution.  The effectiveness of the flavor
determination and the vertex resolution function are both determined
by using a large sample of exclusively reconstructed decays
into flavor eigenstates.

\subsubsection{\it Data Samples}

There are several decay modes which are used to amass a sample of $b
\to c\bar{c}s$ events, which comprise our CP data sample. The
$c\bar{c}$ quarks make either a $\jpsi$, $\psi(2S)$, or $\chic1$.  For
the $\jpsi$ we use only the decay into electron or muon pairs.  The
$\psi(2S)$ decay into lepton pairs and that into $\jpsi \pip \pim$ are
used.  The $\chic1$ decay into $\jpsi \gamma$ is used.  The $s$ quark
hadronizes into a $\KS$, $\KL$, or a $\Kstar$.  Of course only the
decays of the $\Kstar$ into neutral kaons can be used.  Both charged
and neutral two pion decays of the $\KS$ are used.  The $\KL$ are
detected directly in the EMC and the IFR.

Reconstructed \jpsi, \psitwos, and \chic1 are combined with
reconstructed \KS, \KL, or \Kstar to make up a \Bz.  The \KL mode has
only one independent kinematic variable since one constraint must be used
to infer the \KL momentum; only the \KL direction is measured in the
IFR. The energy substituted $B$ mass, $\mes = \sqrt{E_{\rm Beam}^{2} -
  p_{B}^{2}}$ and $\Delta E = E_{\Bz} - E_{\rm Beam}$ are shown in
Figure~\ref{fig:cpsample} for a sample of $33 \times 10^{6}$ \BB
events taken during 1999-2001. 
% The number of events in each decay
%mode, and the corresponding background levels are shown in
%Table~\ref{table:cpsample}.  
There was a significant improvement in the
reconstruction efficiency of the 2001 data compared to that of the
2000 year data, due mostly to improved DCH tracking reconstruction for
$\KS \to \pip \pim$.  This improvement will be applied to the data
from 1999-2000 when that data set is eventually re-analyzed.
\begin{figure}[thb]
  \begin{center}
  \includegraphics[height=0.5\textheight]{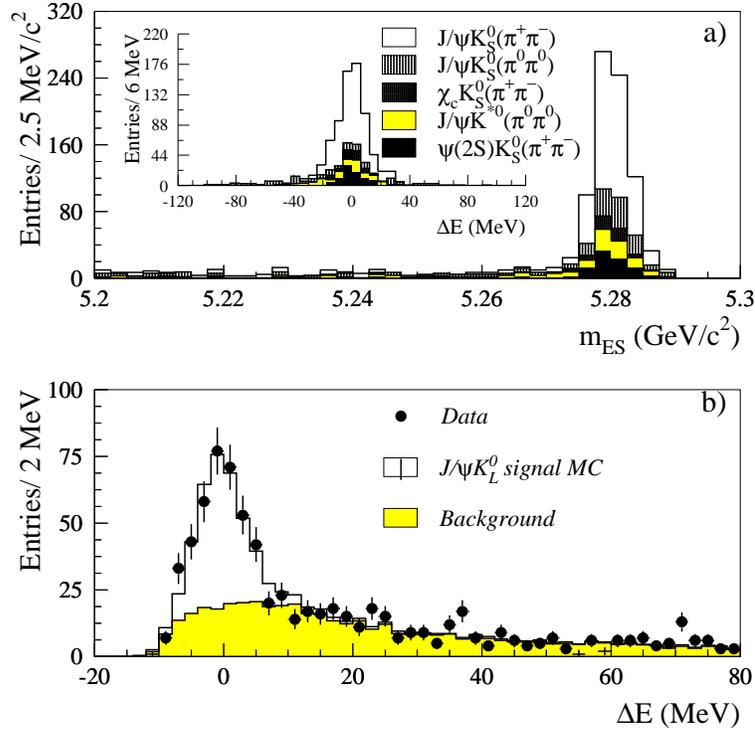}
  \end{center}
  \caption{\it
    a) $\mes$ and $\Delta E$ distributions for the CP=-1 data sample
    (the \bpsikst sample is a mix of CP=-1 and CP=+1, but it is used
as a CP=-1 state with an extra dilution factor of 0.7), 
and b) $\Delta E$ for the CP=+1 sample.
    \label{fig:cpsample} }
\end{figure}

\subsubsection{\it B Flavor Tagging}

The flavor of the other $\Bz$ is determined from the charge of high
momentum leptons, kaons, and slow pions from $\Dstar$ decays.
Electrons are identified using the ratio of EMC energy and track
momentum, the shower shape in the EMC, the $dE/dx$ in the DCH, and the
cherenkov angle in the DIRC.  The efficiency and mis-identification
rate for electrons are shown in Figure~\ref{fig:pid}a. Muons are
identified using the hits in the IFR, and by comparing the DCH track's
extrapolation through the IFR with the IFR hits.  The efficiency and
mis-identification rate from pions are shown in Figure~\ref{fig:pid}b.
The mis-identification rate is somewhat higher than desired due mostly
to the thickness, 4.5 to 6.0 interaction lengths, of the detector, and
to inefficiencies in the RPCs. Charged kaons are identified using the
$dE/dx$ in the SVT and DCH, the cherenkov angle and number of
cherenkov photons in the DIRC.  There is good separation between kaons
and pions up to around 0.6~GeV from dE/dx and the cherenkov threshold
in the DIRC is approximately 0.6~GeV, so there is good Kaon
identification for all momentum.  The Kaon efficiency and
mis-identification for pions are shown in Figure~\ref{fig:pid}c.

\begin{figure}[thb]
  \includegraphics[width=0.3\textwidth]{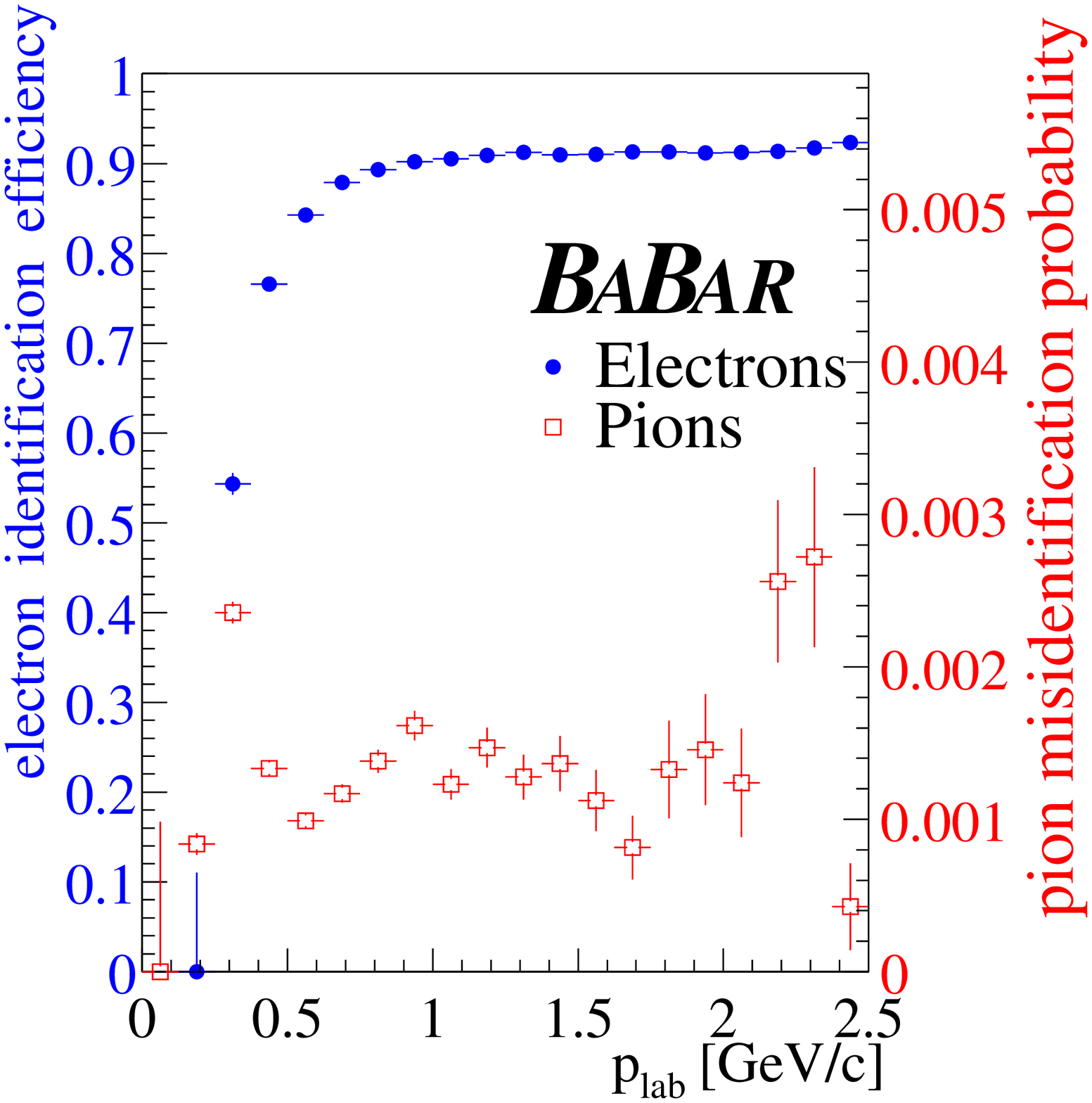}
  \includegraphics[width=0.3\textwidth]{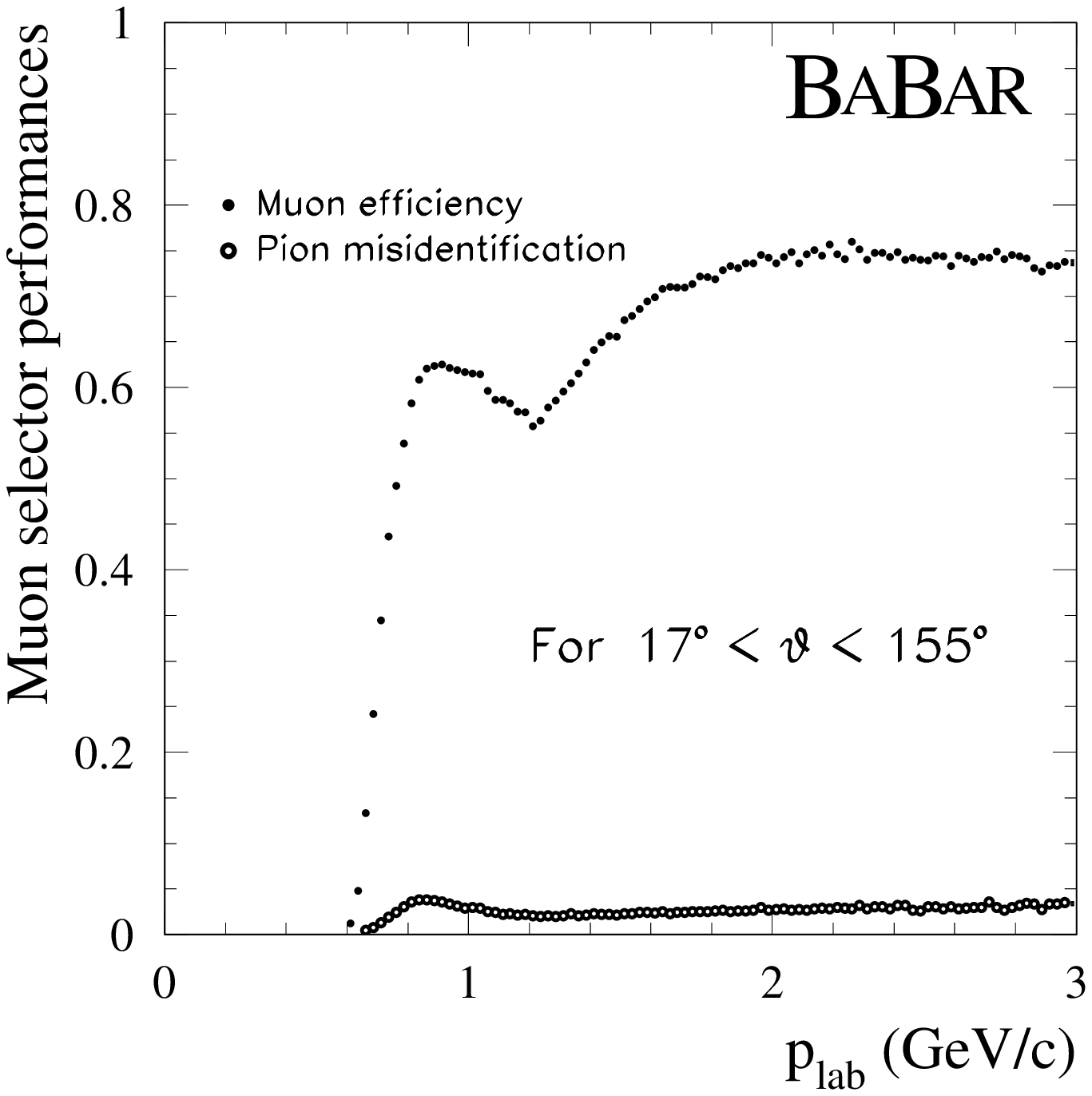}
  \includegraphics[width=0.35\textwidth]{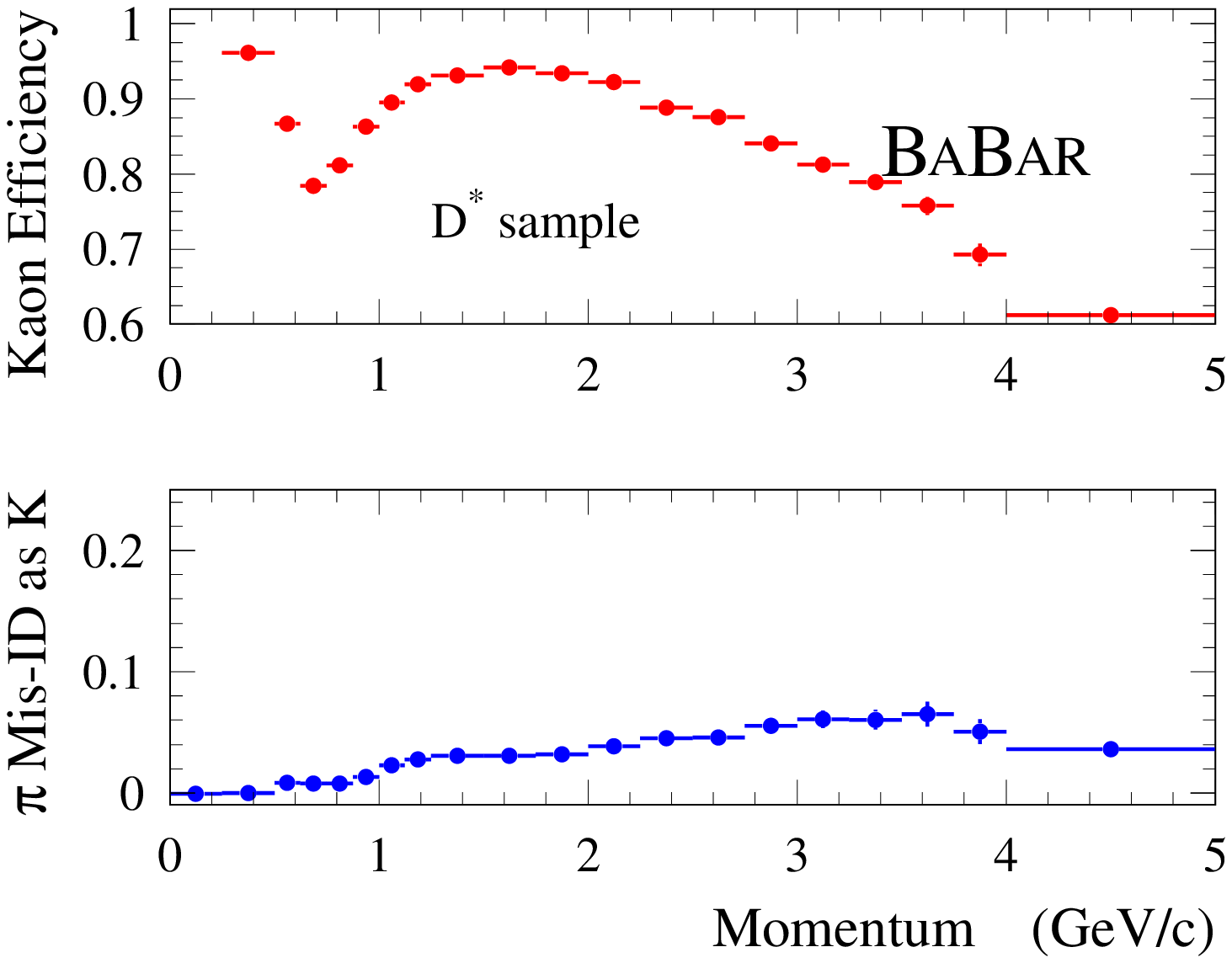}
  \caption{\it
    Particle identification at \babar\ a) electrons,  b) muons, c)
    kaons. The efficiency and mis-identification probability is shown
    as a function of momentum.
    \label{fig:pid} }
\end{figure}

To determine the flavor of the \Bz, each event is categorized,
or tagged,
according to its particle content.  First, events with electrons with
$p_{CM} > 1.0~GeV$ and muons with $p_{CM} > 1.1~GeV$ are used in the
{\it Lepton} category. Next, events with one or more charged kaons are
used in the {\it Kaon} category.  For multiple kaons, the sum of Kaon
charge is used.  An artificial neural network is used for all
remaining events.  The neural network uses slow pions from $\Dstar$
decays and any remaining leptons or charged kaons which fail their
respective selection.  Slow pions are identified by their $p_{CM}$
and the angle between the pion and the thrust axis of all tracks and
neutral EMC clusters from the $\Bz$.  This thrust axis is generally
aligned with the original $\Dstar$ direction, as is the slow pion
direction. Additional leptons are identified using the lepton momentum
and the lepton's isolation with respect to other tracks and clusters
from the other $\Bz$. Two categories are defined from the output of
the neural network, {\it NT1} and {\it NT2}, corresponding to more
certain and less certain events.

The efficiency and mistagging rates are determined from the \Bz mixing
sample in which one $\Bz$ is fully reconstructed in a flavor
eigenstate.  By using data to measure the efficiency and mistag rates,
most systematic effects are canceled. A sample of decays $\Bz \to
\Dstar n\pi$, $\Bz \to D n\pi$, and $\Bz \to \jpsi \Kstar, \Kstar \to
\Kpm \pi$ are used for the flavor eigenstate sample; the \mes for this
sample is shown in Figure~\ref{fig:mixing}.  The tagging efficiency
and mistag rates are shown in Table~\ref{tab:tagging}, along with the
quality factor $Q = \epsilon (1-2\omega)^{2}$.  The statistical power
of the asymmetry measurement is given by $\sigma_{\rm asym} =
\sigma_{0}/\sqrt{(NQ)}$.
\begin{figure}[thb]
  \includegraphics[width=0.95\textwidth]{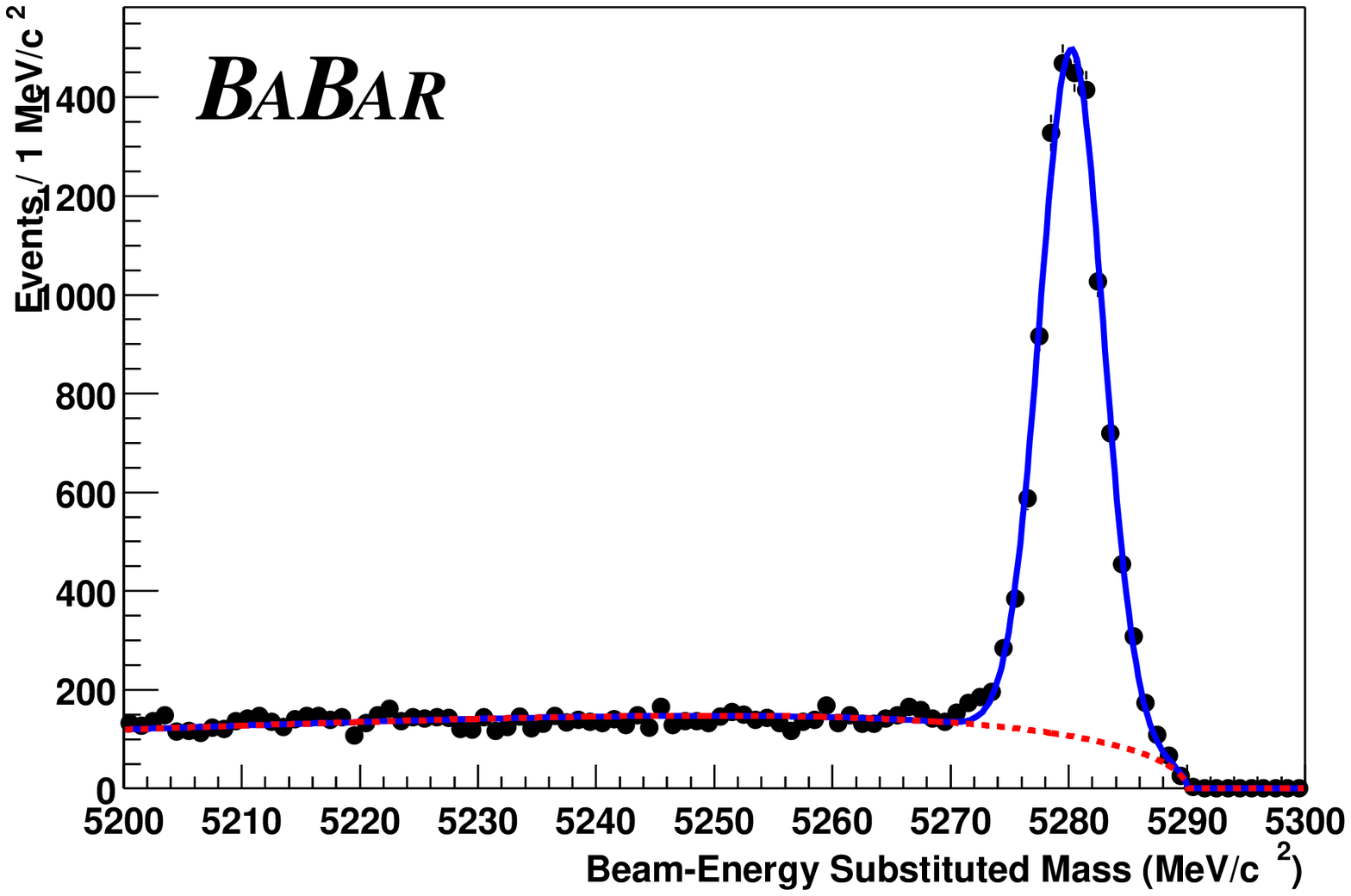}
  \caption{\it
    \mes for the exclusively reconstructed \Bz sample.
    \label{fig:mixing} }
\end{figure} 

\begin{table}[t]  
  \centering
  \caption{ \it Tagging efficiency, mistag rate, and quality factor
    $Q$ from the flavor eigenstate sample. }
  \vskip 0.1 in
  \begin{tabular}{lccc} \hline 
   Category & $\epsilon(\%)$ & $\omega(\%)$ & $Q(\%)$ \\ \hline \hline
   \hline
  Lepton & $10.9 \pm 0.3 $ & $ 8.9 \pm 1.3 $ & $ 7.4 \pm 0.5$ \\
  Kaon & $35.8 \pm 0.5 $ & $ 17.6 \pm 1.0 $ & $ 15.0 \pm 0.9$ \\
  NT1 & $7.8 \pm 0.3 $ & $ 22.0 \pm 2.1 $ & $ 2.5 \pm 0.4$ \\
  NT2 & $13.8 \pm 0.3 $ & $ 35.1 \pm 1.9 $ & $ 1.2 \pm 0.3$ \\ \hline
  Total & $68.4 \pm 0.7 $ &  & $ 26.1 \pm 1.2$ \\ \hline
  \end{tabular}  \label{tab:tagging}
\end{table}

\subsubsection{\it B Vertex and Decay Time}
 
The decay time difference $\deltat$ is measured using the SVT to
determine the $\deltaz$ between the two $\Bz$ decays. We convert 
according to $\deltat = \deltaz/(c\beta\gamma)$ with a correction for
the deviation of the \B flight direction from the beam axis. The vertex of
the exclusively reconstructed CP or Flavor decay is determined by a
constrained fit of the measured tracks, taking into account the
presence of intermediate decay particles such as $D^{0}$ and $\KS$.
For the other $\Bz$, which is not fully reconstructed, the vertex
finding is complicated by the fact that long-lived intermediate
particles are not identified, and the vertex is susceptible to a
bias from $D^{0}$ or $\KS$ decays.  To avoid such bias, tracks which
contribution too much to the vertex $\chi^{2}$ are iteratively removed
from the vertex fit.  Also, to further improve the vertex
measurements, and to allow the use of events in which there is only
one charged track, the $\deltaz$ is determined
using a constraint derived from the exclusive $\Bz$ direction and the
beam spot.  The overall $\deltaz$ resolution is $190 \mum$, with a
core resolution of $110\mu m$ comprising roughly 65\% of the events.

The $\deltat$ resolution function is parametrized as the sum of three
Gaussian distributions.  Roughly speaking the first Gaussian is for
the core of the distribution, the second for the multiple scattering
tail, and the third is nearly flat to account for mis-measured events.
As part of the parametrization, the estimated error for each event
from the $\deltaz$ fit
%, as shown in Figure~\ref{fig:sigzerr}, 
is used.  Decays with higher track multiplicity and higher \pt
tracks have smaller $\sigma_{\deltaz}$.  The resolution function
parametrization is given by
\begin{equation}
 \mathcal{R}(\deltat) = \sum_{\mathrm{core,tail}}
                   \frac{1}{\sqrt{2\pi}S_{i} \sigma_{\deltat}}
                   e^{- \frac{(\deltat - m_{i,c}\sigma_{\deltat} )^{2}}
                      { 2 (S_{i}\sigma_{\deltat})^{2} } } + 
                   \frac{1}{\sqrt{2\pi} \sigma_{\mathrm{outlier}} }
                   e^{- \frac{(\deltat)^{2}}
                      { 2 (\sigma_{\mathrm{outlier}})^{2} } }
\end{equation}
where the estimated error for each event, $\sigma_{\deltat}$, are
scaled by free parameters $S_{i}$, and an offset to account for
remaining bias from charm meson decays is parametrized with a linear
dependence on the estimated error.  The sample of exclusively
reconstructed \Bz is used, in the combined fit described below, to
determine the values of the free parameters.
%, as listed in Table~\ref{tab:vertex}.

\subsection{CP Violation Results}

The CP violating asymmetry is measured using an unbinned maximum
likelihood fit. The likelihood function for signal events is simply
the expression given for the decay rate.  To easily include the
statistical uncertainty from  the flavor tagging and vertex
resolution parametrization, we use a combined fit to both the CP
events and the Flavor events.

In addition,
to help avoid experimenters bias, the CP fit is done blind to the
value of $\stwob$.  Prior to finalizing the measurement, the asymmetry
parameter used in the fit was 
\begin{equation}
a^{\mathrm{Hidden}} = 
               \left\{ 
                  \begin{array}{c}1\\-1\end{array}
               \right\} 
               \times  a  + C   
\end{equation}
where the value of $C$ and the choice of 1 or -1 was fixed, arbitrary
and hidden.  To also hide any visual asymmetry, the $\deltat$
distribution was altered in plotting by $\deltat^{\mathrm{Hidden}} =
\mathrm{S}_{\rm Tag} \deltat + \mathrm{Offset}$, where again the
$\mathrm{Offset}$ was fixed, arbitrary and hidden.  The asymmetry result
was hidden until the analysis was essentially complete.

The $\deltat$ distribution for $\Bz$ and $\Bzb$ tags is shown in
Figure~\ref{fig:cpdeltat} for the $CP=-1$ decays modes $\bpsiks$, $\bpsitwosks$,
and $\bchiconeks$ and the $CP=+1$ mode $\bpsikl$.  Also shown is the raw asymmetry as a function of the
$\deltat$. The $\sin$ oscillation is readily apparent.  
%The same
%distributions are shown for the $\bpsikl$ decay mode in
%Figure~\ref{fig:cpdeltatklong}, with the contribution from backgrounds
%shown separately. 
We find a value for \stwob of\cite{babarstwob} 
\begin{equation}
  \stwob = 0.59 \pm 0.14(\textrm{stat}) \pm 0.05 (\textrm{syst})
\end{equation}
The systematic error is largely due to uncertainties in the $\deltaz$
determination and the mistagging probability, with smaller
contributions from uncertainty in the values of $\deltamd$ and
$\tau_{\Bz}$.  The $\bpsikl$ has an additional systematic contribution
from its larger background.  
\begin{figure}[thb]
  \includegraphics[height=0.5\textheight]{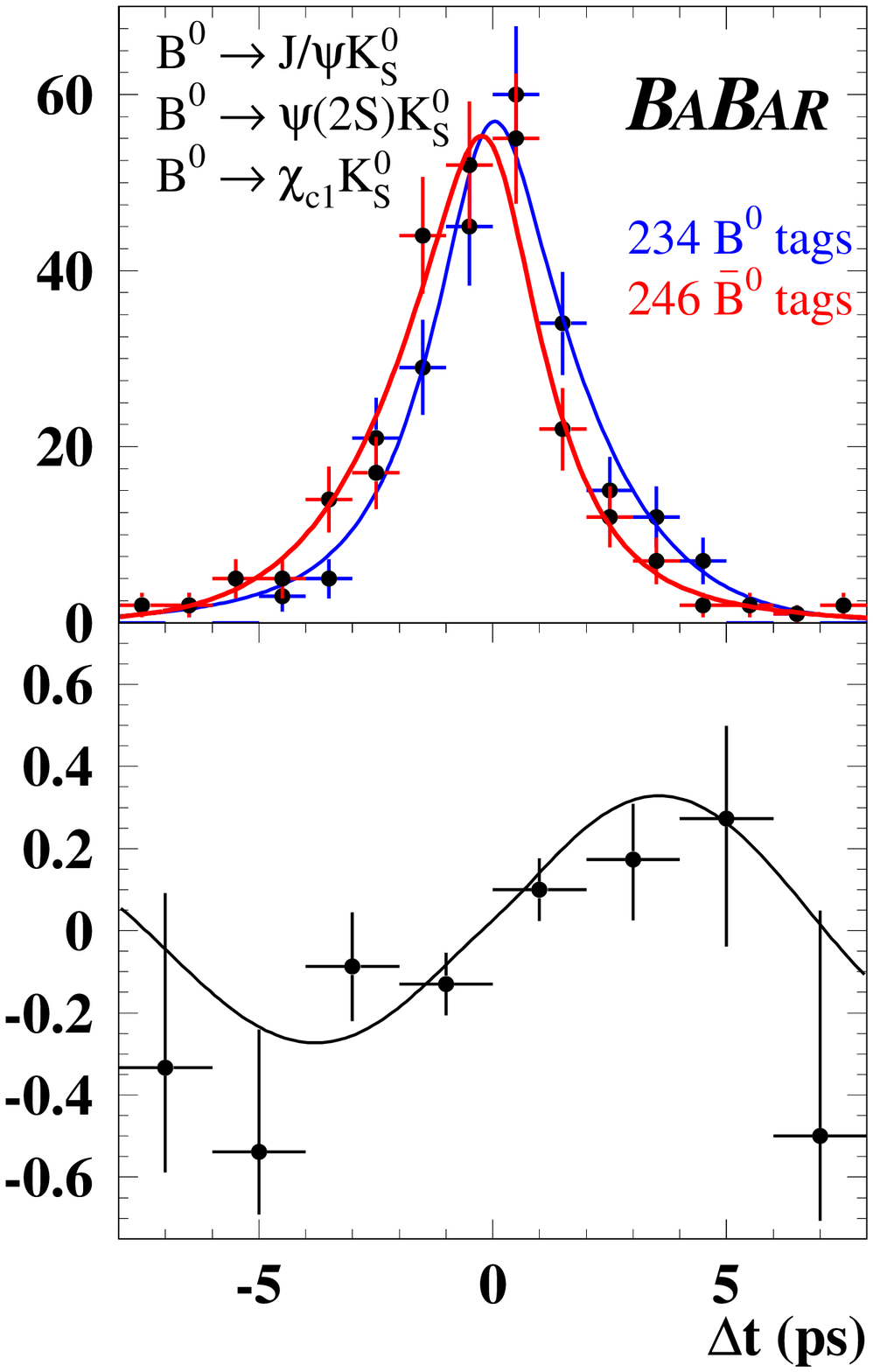}
  \includegraphics[height=0.5\textheight]{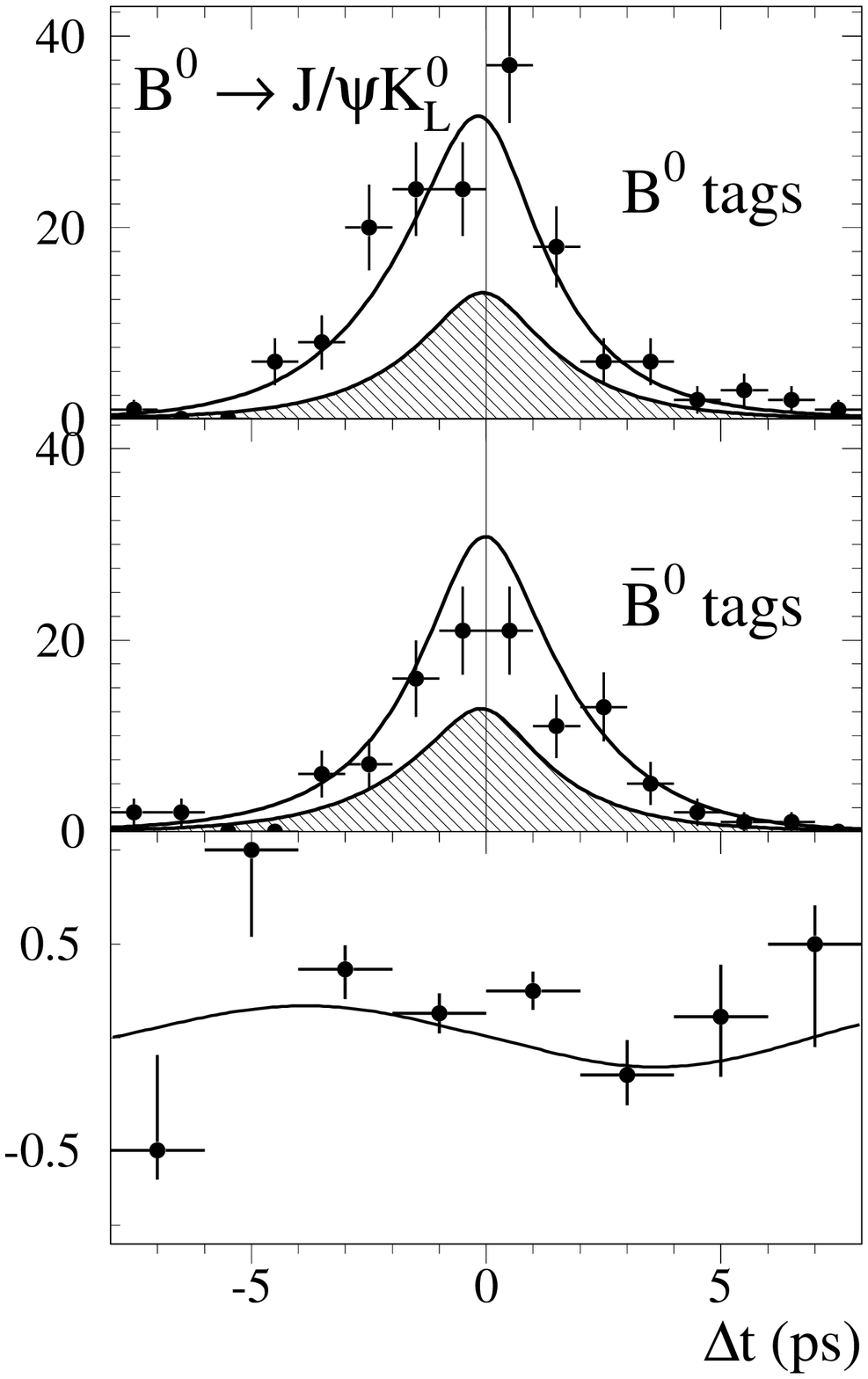}
  \caption{\it The \deltat distribution for \Bz and
    \Bzb tags and the Asymmetry, $\mathcal{A}(\deltat) = (N_{\Bz} -
    N_{\Bzb})/(N_{\Bz} + N_{\Bzb})$  for a) the CP=-1 sample (\KS decays), and b)
    the CP=+1 sample(\KL decays).  
    \label{fig:cpdeltat} }
\end{figure}

Our measured value of \stwob and the constraints from other
measurements are shown in Figure~\ref{fig:triangle}
\begin{figure}[thb]
  \begin{center}
  \includegraphics[height=0.5\textheight]{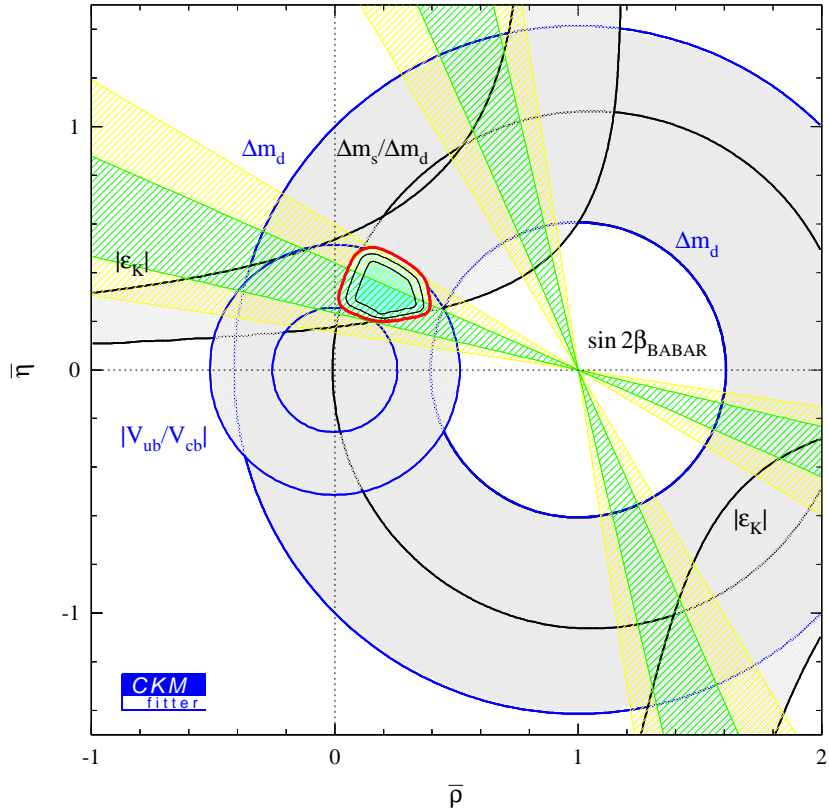}
  \end{center}
  \caption{\it
    \babar measurement of \stwob in the CKM $\rho-\eta$ plane, along
    with constraints from other measurements.
    \label{fig:triangle} }
\end{figure}
At the level of the statistical uncertainty in \stwob  there is good
agreement among the measurements in the $\rho-\eta$ plane.  As the
measurement of \stwob improves statistically, the comparison will
depend heavily on the theoretical uncertainties in translating
measurements of $b \to u l \nu$, \deltamd, and $\epsilon_{K}$ into CKM
parameters.  When \deltams mixing is measured, the theoretical
uncertainty will decrease by using the ratio $\deltams/\deltamd$, and
improvement is possible in \Vub given more data and expected
improvements in lattice calculations.  However, a test of the entire CKM
picture at the 5\% level will likely require an accurate and theoretically
clean measurement of one of the other CP violating angles.

\section{B Lifetime and Mixing}

A measurement of the fundamental parameters of the $B$ meson system,
the $\Bz$ and $\Bpm$ lifetimes, and the \Bz mixing parameter \deltamd,
provide crucial input to the CP asymmetry measurements and to the
constraints on the parameters of the CKM matrix.

\subsection{B Lifetime Measurement} 

The $\Bz$ and $\Bpm$ lifetimes are measured with a sample of
exclusively reconstructed decays.
The $\deltat$ distribution of the  6967 $\Bz$
and 7266 $\Bpm$ events is shown in Figure~\ref{fig:blifetime}.  The
resolution function is parametrized in a somewhat different way than
in the CP violation measurement, using a Gaussian+Exponential
convolution. As in the asymmetry measurement, an unbinned maximum
likelihood fit is performed. The
results for the lifetimes\cite{babarlife} are listed in Table~\ref{tab:results}.
where the systematic uncertainty comes from the {\it outlier}
contribution to the resolution function (0.011), resolution
parameterization (0.011), and absolute Z scale (0.008). Systematic
for the lifetime ratio come from differences in the resolution
function for $\Bz$ and $\Bpm$ (0.006), and {\it outliers} (0.005).
\begin{figure}[thb]
  \includegraphics[width=0.45\textwidth]{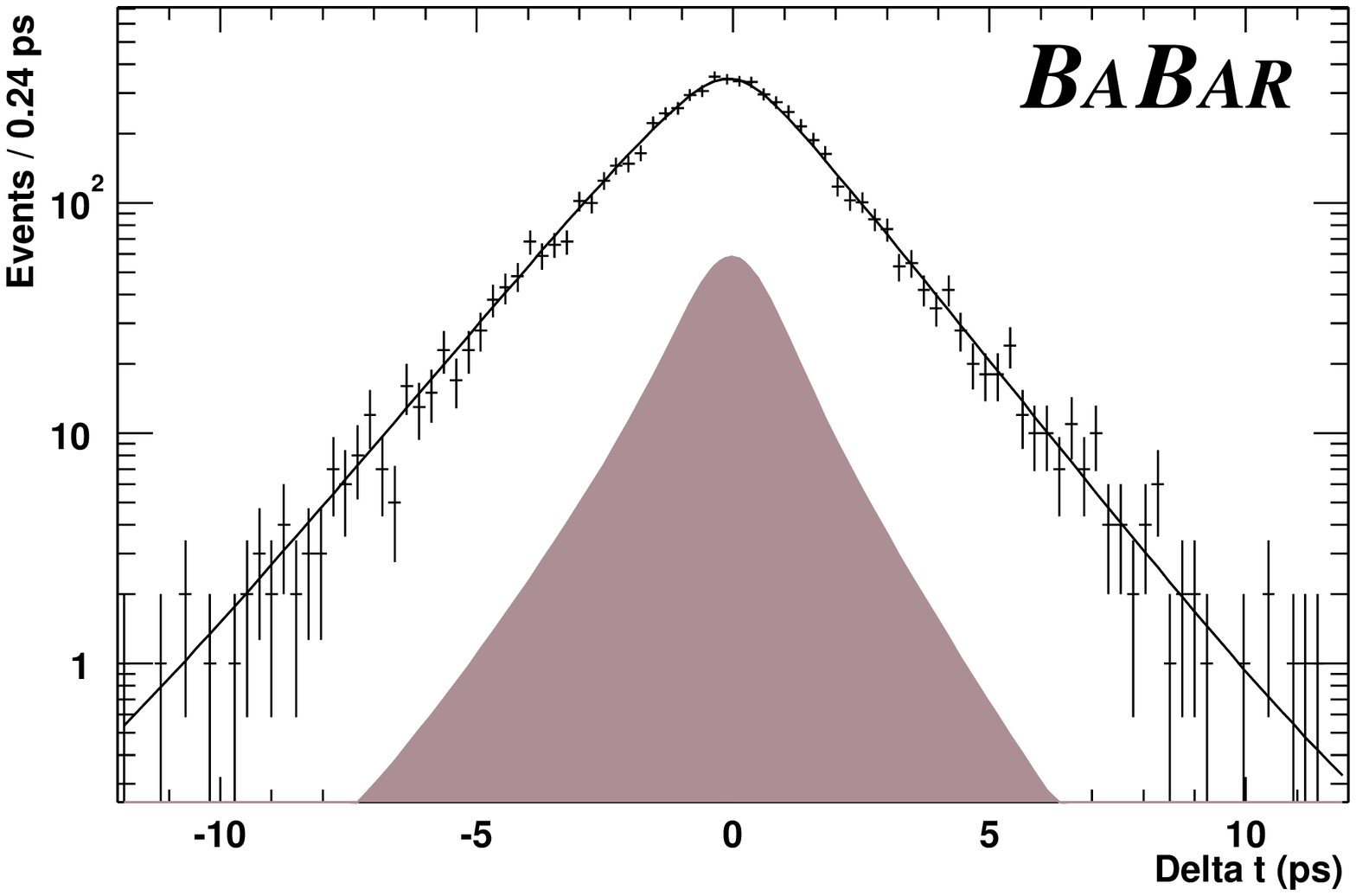}
  \includegraphics[width=0.45\textwidth]{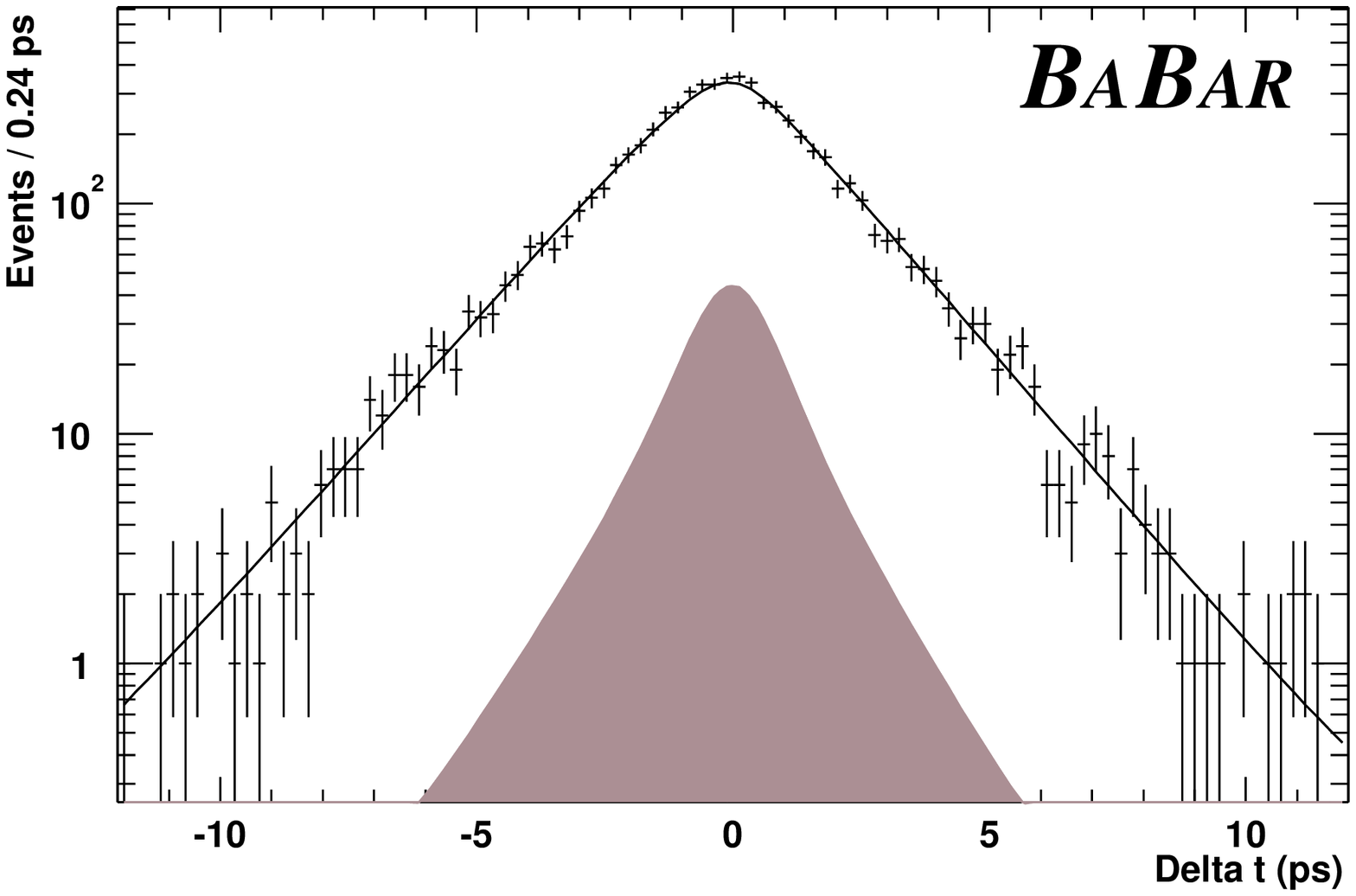}
  \caption{\it
    The \deltat distributions for a) \Bz and b) \Bpm decays, along
    with the fit for the lifetime.   
    \label{fig:blifetime} }
\end{figure}
The ratio $\tau_{\pm}/\tau_{0}$ is sensitive to the \B decay constant,
$f_{\B}$, due to the contribution from W-exchange diagrams which occur
for \Bz decay only.

\begin{table}[thb]
  \centering
  \caption{ \it Results from the \babar Experiment. }
  \vskip 0.1 in
  \begin{tabular}{|lc|} \hline
    Quantity & Result \\
    \hline
    \hline
    \hline
    $\tau_{\Bz} $ &  $1.546 \pm 0.032 ({\rm stat}) \pm 0.022 ({\rm
      syst}) \ps$ \\
    $\tau_{\Bpm} $ & $ 1.673 \pm 0.032 ({\rm stat}) \pm 0.022 ({\rm
      syst})  \ps$ \\
    $\tau_{\Bpm}/\tau_{\Bz} $ & $ 1.082 \pm 0.026 ({\rm stat}) \pm 0.012 ({\rm
syst}) $ \\  \hline \hline
    $\deltamd$ Exclusive & $ 0.519 \pm 0.020 (\mathrm{stat}) \pm 0.016 (\mathrm{syst}) \ps^{-1}$ \\
    $\deltamd$ dilepton &  $0.499 \pm 0.010 (\mathrm{stat}) \pm 0.012
    (\mathrm{syst}) \ps^{-1}$ \\ \hline \hline
    $ \mathcal{B}(\Bz \to \Kstarz \gamma)$ & $(4.39 \pm 0.42 (\mathrm{stat})
\pm 0.27 (\mathrm{syst})) \times 10^{-5}$ \\
    $A_{CP}$ $\Kstarz \gamma$  & $  -0.035 \pm
    0.094(\mathrm{stat}) \pm 0.022(\mathrm{syst}) $ \\ \hline \hline
    $ \mathcal{B}(\Bztopipi)$ & $ (4.1 \pm 1.0 (\mathrm{stat})
\pm 0.7 (\mathrm{syst})) \times 10^{-6} $ \\
    $ \mathcal{B}(\Bztokpi)$ & $ (16.7 \pm 1.6 (\mathrm{stat})
\pm^{+1.2}_{-1.7} (\mathrm{syst})) \times 10^{-6} $ \\
    $ \mathcal{B}(\Bz \to \Kpm \Kmp)$ & $< 2.5 \times 10^{-6}$ \\
    $A_{CP}$ $K \pi$  &  $ -0.19 \pm
    0.10(\mathrm{stat}) \pm 0.03(\mathrm{syst}) $\\ \hline \hline  
    $ \mathcal{B}(\Bpm \to \phi \Kpm) $ & $(7.7^{+1.6} 
_{-1.4 } (\mathrm{stat}) \pm 0.8(\mathrm{syst}) ) \times 10^{-6} $ \\ 
    $ \mathcal{B}(\Bz \to \phi \KS) $ & $(8.1^{+3.1}
_{-2.5 } (\mathrm{stat}) \pm 0.8(\mathrm{syst}) ) \times 10^{-6} $ \\ 
    $ \mathcal{B}(\Bpm \to \phi \Kstarpm) $ & $(9.6^{+4.1} 
_{-3.3 } (\mathrm{stat}) \pm 1.7(\mathrm{syst}) ) \times 10^{-6} $ \\ 
    $ \mathcal{B}(\Bz \to \phi \Kstarz) $ & $(8.6^{+2.8}
_{-2.4 } (\mathrm{stat}) \pm 1.1(\mathrm{syst}) )  \times 10^{-6} $ \\ \hline 
  \end{tabular}
  \label{tab:results}
\end{table}

\subsection{B Mixing Measurement}

The $\Bz$ mixing parameter $\deltamd$ has been measured using two
different samples.  The first is the exclusively reconstructed sample
used in the CP asymmetry measurement, and the second is a sample of
events in which both $\Bz$ decay semi-leptonically.

\subsubsection{Mixing in Exclusive Decays}

The  $\Bz$ mixing parameter $\deltamd$ is measured using the
exclusively reconstructed sample and the inclusive flavor
tagging of the other $\Bz$ to determine the fraction of mixed $\Bz\Bz$
or $\Bzb\Bzb$ events.  The asymmetry between mixed and un-mixed
events is given by
 \begin{equation} \mathcal{A}_{\rm mixing}(\deltat) = (\textrm{Mixed}-
    \textrm{Un-Mixed})/(\textrm{Mixed} +  \textrm{Un-Mixed}) =
  (1-2\omega)\cos{(\deltamd \deltat)}
 \end{equation}
As in the CP measurement, the asymmetry is modified by incorrect
flavor tagging and vertex resolution and  the value of $\deltamd$ is
again found from an unbinned maximum likelihood fit. The $\deltat$
distribution is shown in Figure~\ref{fig:mixingcos} along with the
asymmetry as a function of  $\deltat$. The preliminary value is listed
in Table~\ref{tab:results}\cite{babarmixlp}.
The systematic uncertainties are due to  corrections taken from
simulation (0.009), the $\deltat$ measurement scale, boost, and
alignment uncertainty (0.008),  $\Bz$
lifetime (0.006),  and backgrounds (0.005). 
\begin{figure}[thb]
  \begin{center}
  \includegraphics[height=0.5\textheight]{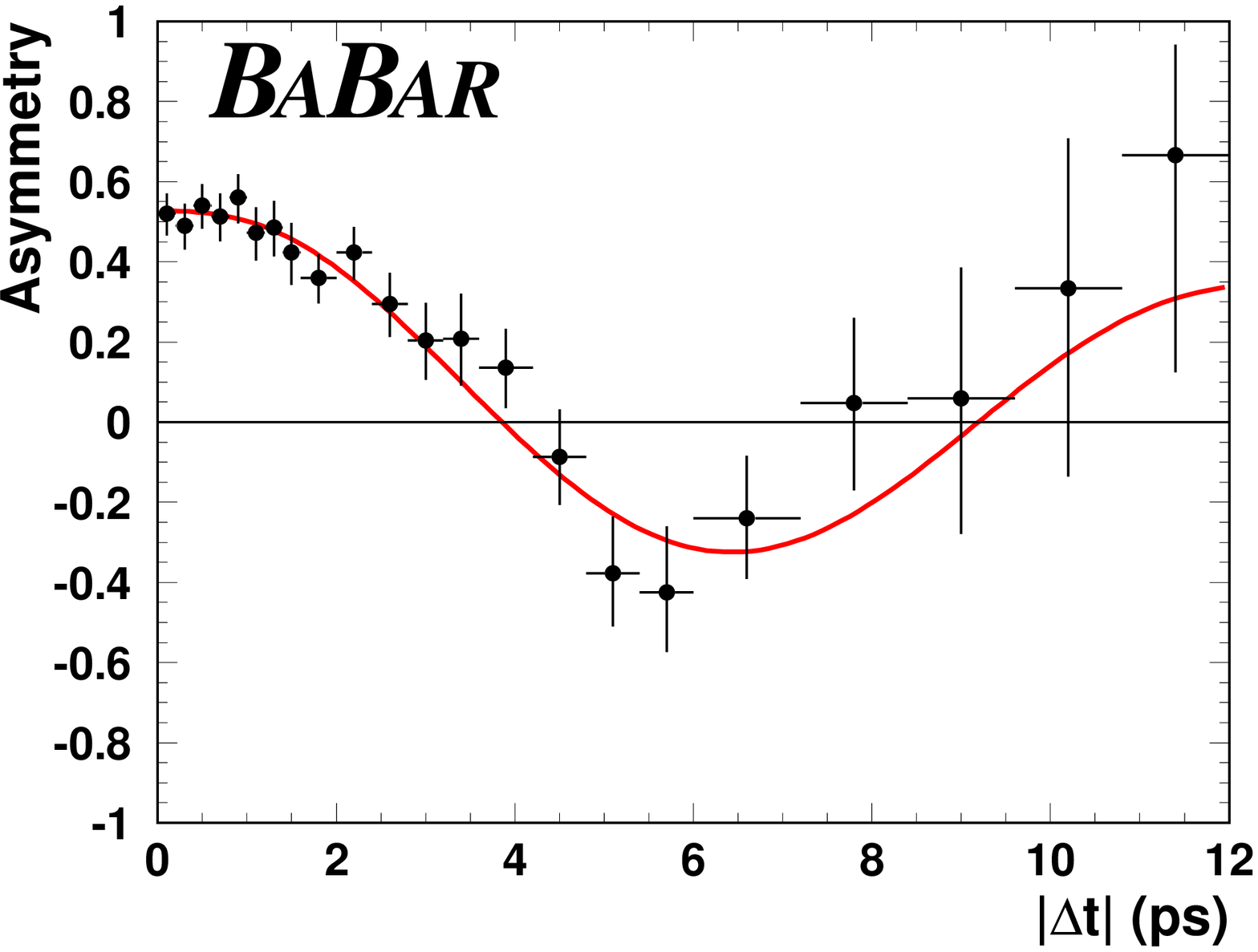}
  \end{center}
  \caption{\it
    The \Bz mixing asymmetry as a function of \deltat, with the fit
    from \deltamd for the exclusively reconstructed sample.
    \label{fig:mixingcos} }
\end{figure}

\subsubsection{Mixing in Di-Lepton Decays}

Alternatively the $\Bz$ mixing parameter can be measured with a sample
of events in which both $\Bz$ decay semileptonically.  Compared with
the exclusively reconstructed sample above, the use of semi-leptonic
decays supplies more events, with greater backgrounds.  In
semi-leptonic events $\Bpm$ decays cannot be fully separated from
$\Bz$ decays, and these extra events dilute the mixing asymmetry. In
addition there are backgrounds from cascade decays, $b\to c
\to l$. The mixing asymmetry is given by
\begin{equation}
 \mathcal{A}(\deltat) = (1-2\omega)
    \frac{ e^{-\Gamma^{0} |\deltat |} \cos{(\deltamd \deltat)} +
      R \frac{\Gamma^{\pm}}{\Gamma^{0}} e^{-\Gamma^{\pm}|\deltat|} }
         { e^{-\Gamma^{0} |\deltat |} +
      R \frac{\Gamma^{\pm}}{\Gamma^{0}} e^{-\Gamma^{\pm}|\deltat|} } 
\end{equation}
where $\omega$ is the mistag Probability and $R =
\frac{b_{+}^{2}f_{\pm}}{b_{0}^{2}f_{00}}$ is the ratio of
semi-leptonic branching ratios and production rates for $\Bz$ and
$\Bpm$.  The ratio $R$ is fit from the data to avoid systematic
uncertainty from the lack of sufficiently accurate measurements.

The mixing asymmetry is shown in Figure~\ref{fig:leptonmixing}. The
preliminary result is listed in Table~\ref{tab:results}.
The systematic uncertainties arise mostly from the $\Delta
z$ resolution function (0.009), cascade backgrounds (0.006) and
dependence on the $\Bz$ and $\Bpm$ lifetimes (0.004).
\begin{figure}[thb]
  \begin{center}
  \includegraphics[height=0.5\textheight]{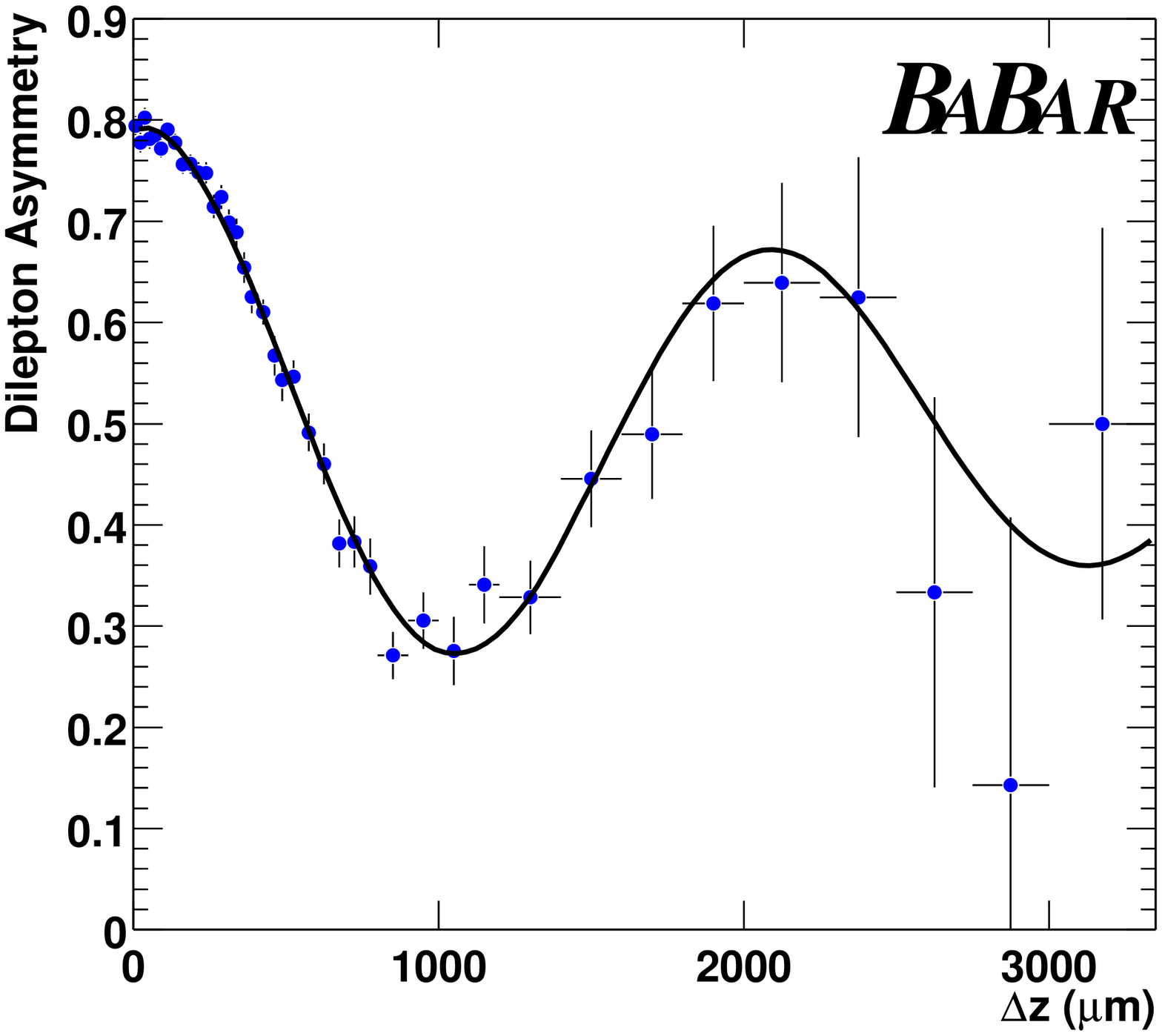}
  \end{center}
  \caption{\it
    The \Bz mixing asymmetry as a function of \deltat, with the fit
    from \deltamd for the di-lepton sample.
    \label{fig:leptonmixing} }
\end{figure}

These measurements of \deltamd have an accuracy comparable to the
previous world average.  When combined with an equally accurate
measurement of \deltams they will provide a theoretically clean
constraint on the CKM matrix.

\section{Rare B Decays}

The bulk of $B$ decays are examples of the quark level process $b
\to c W$, and the CP asymmetry and $\Bz$ mixing measurements
described above use decays of this type.  However, to measure the
unitarity triangle angles $\alpha$ or $\gamma$ will require
measurements of the Cabibbo suppressed decay $b \to u W$.
Also rare decays involving loop diagrams, commonly called penguin diagrams,
offer the potential to detect the influence of non-Standard Model
physics on $B$ decays.  Here we present several \babar branching fraction
and direct CP asymmetry measurements
for such rare decays.

As in any rare decay, the challenge in measuring rare \B meson decay
modes lies in separating signal from background.  For decays which are
exclusively reconstructed, there are two kinematic variables which are
effective in separating \B decays from backgrounds.  At \babar we
typically use the \B mass and energy as our kinematic variables.  A
pair of nearly independent variables are:
    \begin{equation}
      \mes = \sqrt{(E_{\rm beam}^{CM})^{2} - (p^{CM}_{\Bz})^{2}}  \hspace*{1cm}
      \Delta E = E_{\Bz}^{CM} - E_{\rm beam}^{CM}.
    \end{equation}
We use the difference in measured energy and beam energy to remove any
variations in the beam energy.  Then  
the beam energy is substituted in the mass
variable \mes to remove most of the correlation between \mes
and $\Delta E$, and to improve its resolution.  The resolution of
\mes is generally dominated by the beam energy spread of order
2.5~MeV, while the detector resolution dominates $\Delta E$.

\subsection{Radiative Penguin Decays}

Decays of the type $B \to \Kstar \gamma$ can only occur through
radiative penguin decays, with the diagram shown in
Figure~\ref{fig:feynman}a.  
\begin{figure}[thb]
  \includegraphics[bb=95 550 325
690,clip=true,width=0.45\textwidth]{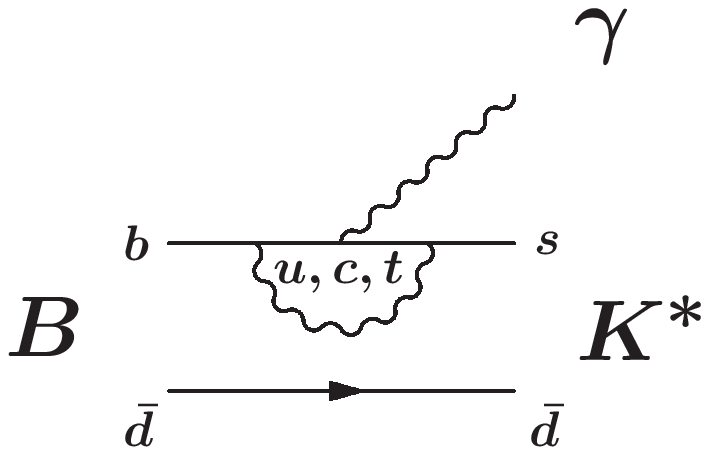}
  \includegraphics[width=0.45\textwidth]{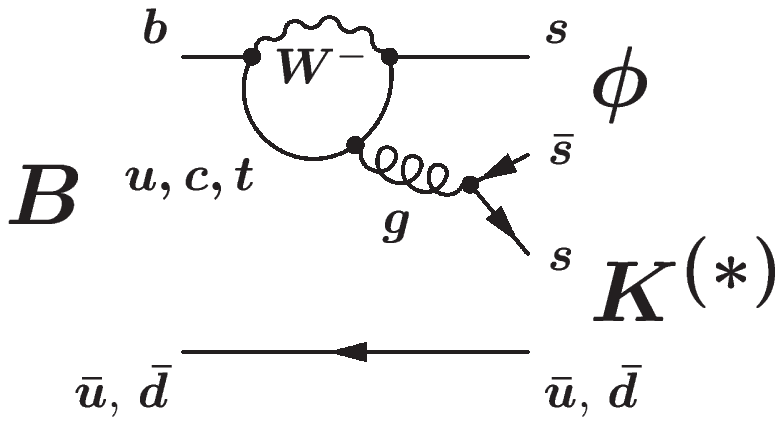}
  \includegraphics[bb=95 550 325 690,clip=true,width=0.45\textwidth]{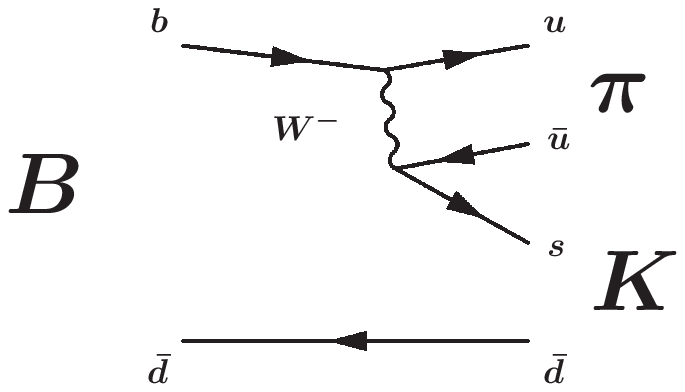}
  \hspace*{1.0cm} \includegraphics[bb=95 550 325
           690,clip=true,width=0.45\textwidth]{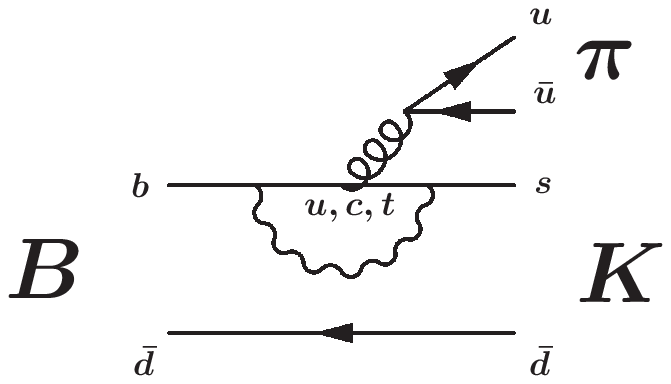}
  \caption{\it
    Feynman diagrams for a) $\B \to \Kstar \gamma$, b) $\B \to \phi
    K$, c) \Bztokpi tree-level, and d) \Bztokpi penguin.
    \label{fig:feynman} }
\end{figure}
Backgrounds to this decay arise from
continuum $q\bar{q}$ events with a leading $\piz$ or $\eta$ and from
events with a high momentum $\gamma$ from initial state radiation.
Event shape variables are effective in separating these backgrounds
from \B decays. The signal for the decay $\Bz \to \Kstarz
\gamma$ and $\Kstarz \to \Kp \pim$ is shown in
Figure~\ref{fig:kstargamma}.  With a signal of $139.2 \pm 13.1$
events, \babar finds the branching fraction\cite{babarkstarg} tabulated in Table~\ref{tab:results}.
The systematics on the asymmetry are due to limits for charged Kaon sign
asymmetries in the detector.  More data will be needed to accurately
search for the small expected CP asymmetries.
\begin{figure}[thb] 
  \includegraphics[width=0.95\textwidth]{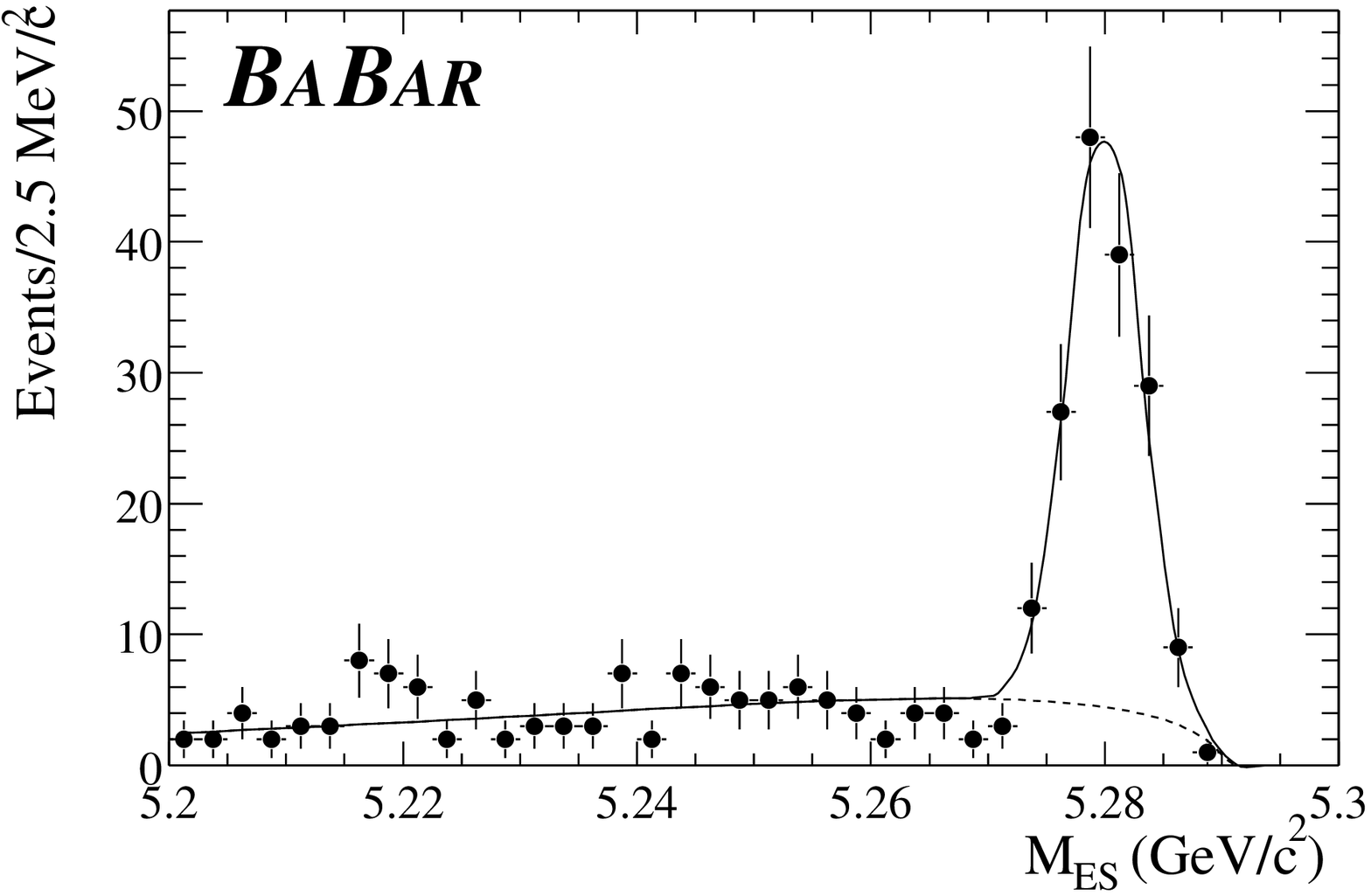}
  \caption{\it
    \mes for the decay $\Bz \to \Kstarz \gamma$.
    \label{fig:kstargamma} }
\end{figure}

\subsection{Charmless Decays}

The angle $\alpha$ of the CKM unitarity triangle can potentially be
measured in the the time dependent asymmetry in $\Bztopipi$.  Unfortunately,
this asymmetry will be difficult to interpret, due to competing
contributions from tree-level and penguin diagrams, shown in
Figure~\ref{fig:feynman}b and c.  However, this
confusion might be resolved with theoretical input, or by measuring
the \Bz and \Bzb rates for the decay $\Bz \to 
\piz \piz$. The first step is to accumulate a sample of $\Bztopipi$ decays.

Observation of $\Bztopipi$ is also complicated by the need to
separate $\pi\pi$ and $K \pi$ decays.  \babar's DIRC was designed to
provide excellent $K/\pi$ separation for these decay modes.  The
DIRC's $K/\pi$ separation can be seen in Figure~\ref{fig:dirckpi} from
a sample of $\Dz \to \Kpm \pimp $ decays.  The DIRC's
response, an event shape variable, and the kinematic variables \mes
and $\Delta E$ are combined in an unbinned maximum likelihood fit to
the rate for both $\Bztopipi$ and $\Bztokpi$.  The branching fraction results\cite{babarpipi},
 are listed in
Table~\ref{tab:results}.  Often it is difficult to visualize the signal
detected using the maximum likelihood technique, since the signal is
identified in a many-dimensional space of variables.  To see as much
of the signal as possible, in a way that shows its separation from
background, we apply an explicit cut on all other variables and show
in  Figure~\ref{fig:bpipi} the signals in the \mes variable.  The
small $\Bztopipi$ branching fraction is consistent with other experiments;
at this level measurements of the time dependent asymmetry will be
possible although accurate measurements will take several year's more data.
\begin{figure}[thb]
    \begin{center}
  \includegraphics[height=0.3\textheight]{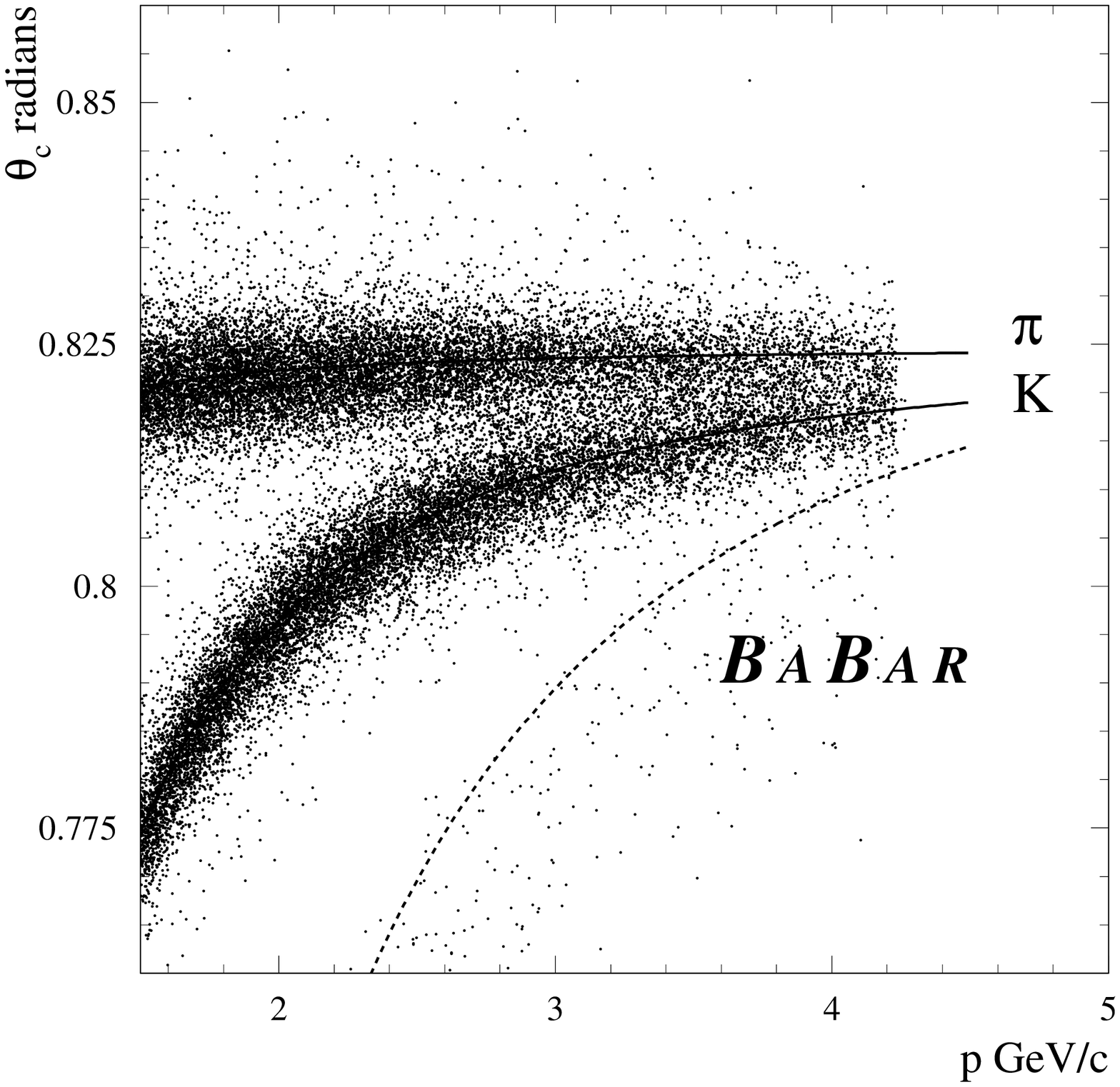}
  \end{center}
  \caption{\it
    The cherenkov angle as a function of momentum measured by the DIRC for a sample of kaons and
    pions from \Dz decays.
    \label{fig:dirckpi} }
\end{figure}

\begin{figure}[htb]
  \begin{center}
  \includegraphics[height=0.45\textheight]{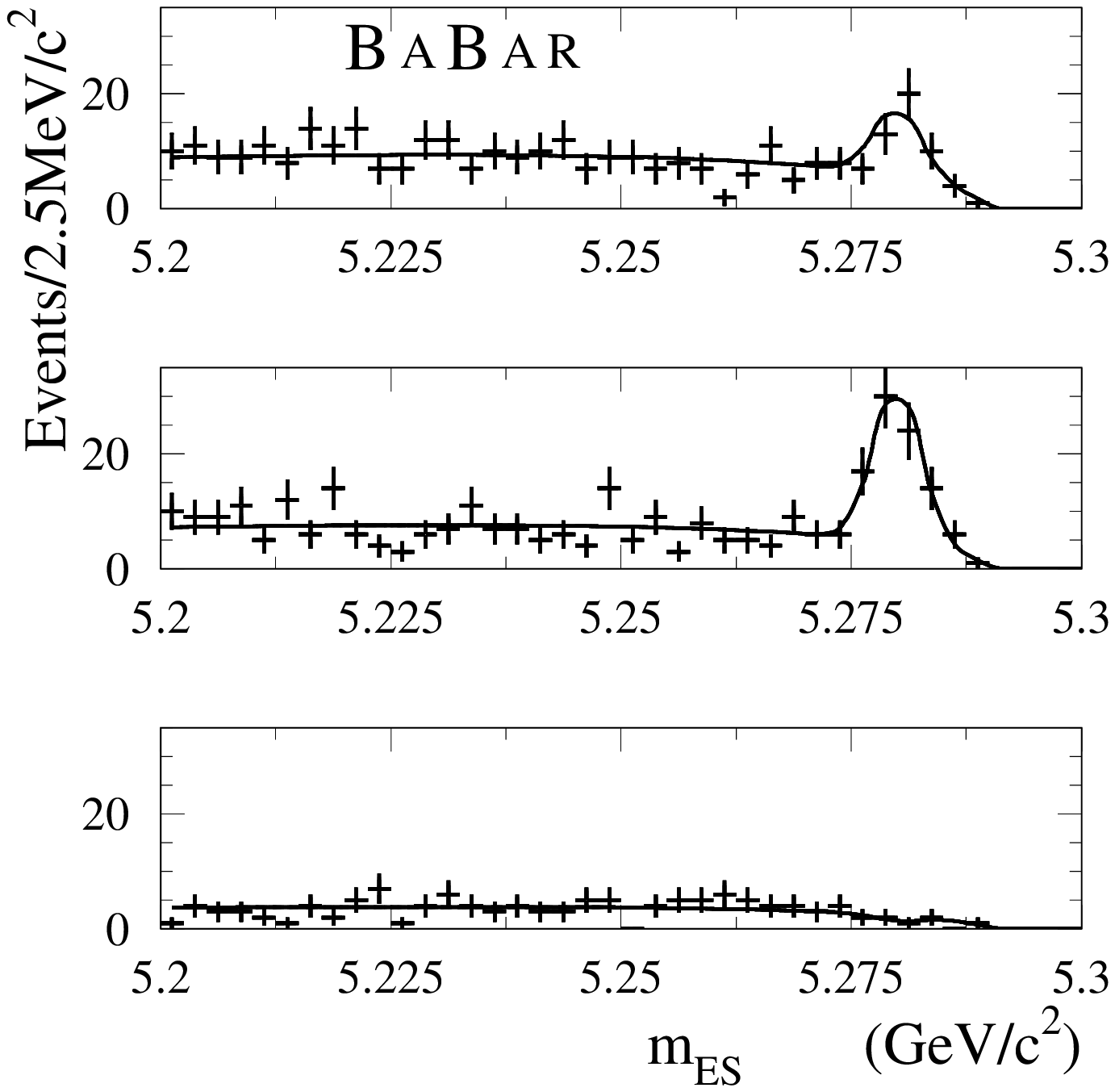}
  \end{center}
  \caption{\it
    The \mes distribution for a) \Bztopipi, b) \Bztokpi, and c) $\Bz
    \to \Kpm \Kmp$.  Discrete cuts have been applied to enhance the
    signal to background rate for each decay mode.
    \label{fig:bpipi} }
\end{figure}

\subsection{Penguin Decays}

Decays of the type $B \to \phi K$, with a quark level decay of
$b \to sss$, can occur only through gluonic penguin diagrams,
as shown in Figure~\ref{fig:feynman}b.
As such these decays are sensitive to the existence of non-Standard
Model particles in the loop which interfere with the $u$,$c$, or $t$
quarks.  

The decays $\bphikp$, $\bphiks$, and $\bphikst$ are measured using a
maximum likelihood fit using an event shape variable, the measured
$\phi \to \Kp \Km$ mass, the helicity angle of the $\phi$ (and
$\Kstar$) decay, and the kinematic variables \mes and $\Delta E$. The
\mes distributions for these decay modes are shown in
Figure~\ref{fig:phik}, and the corresponding branching fractions\cite{babarphik} are
listed in Table~\ref{tab:results}.  The eventual measurement of the
\begin{figure}[htb]  
 \begin{center}
 \includegraphics[height=0.25\textheight]{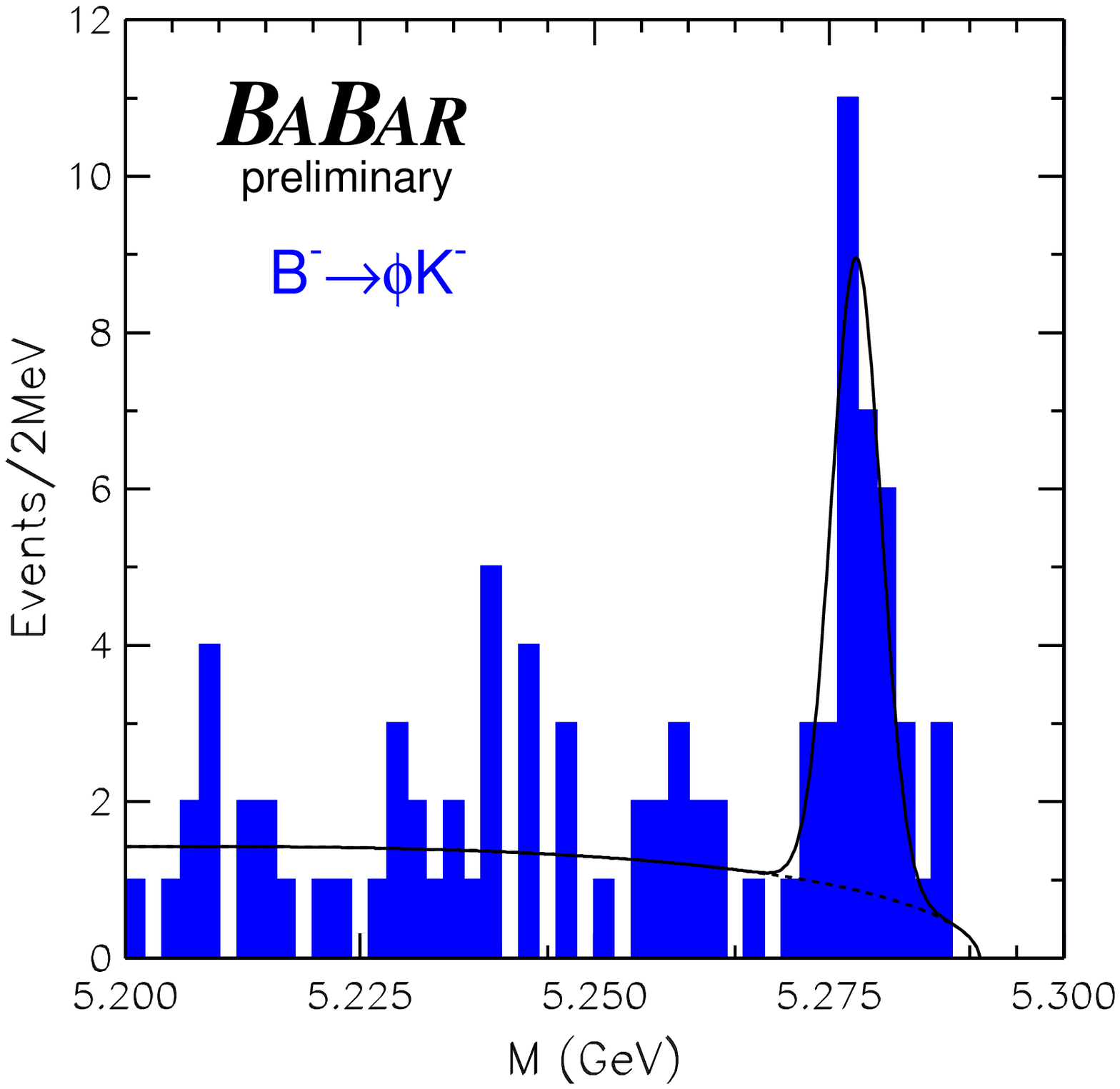}
 \hfill
 \includegraphics[height=0.25\textheight]{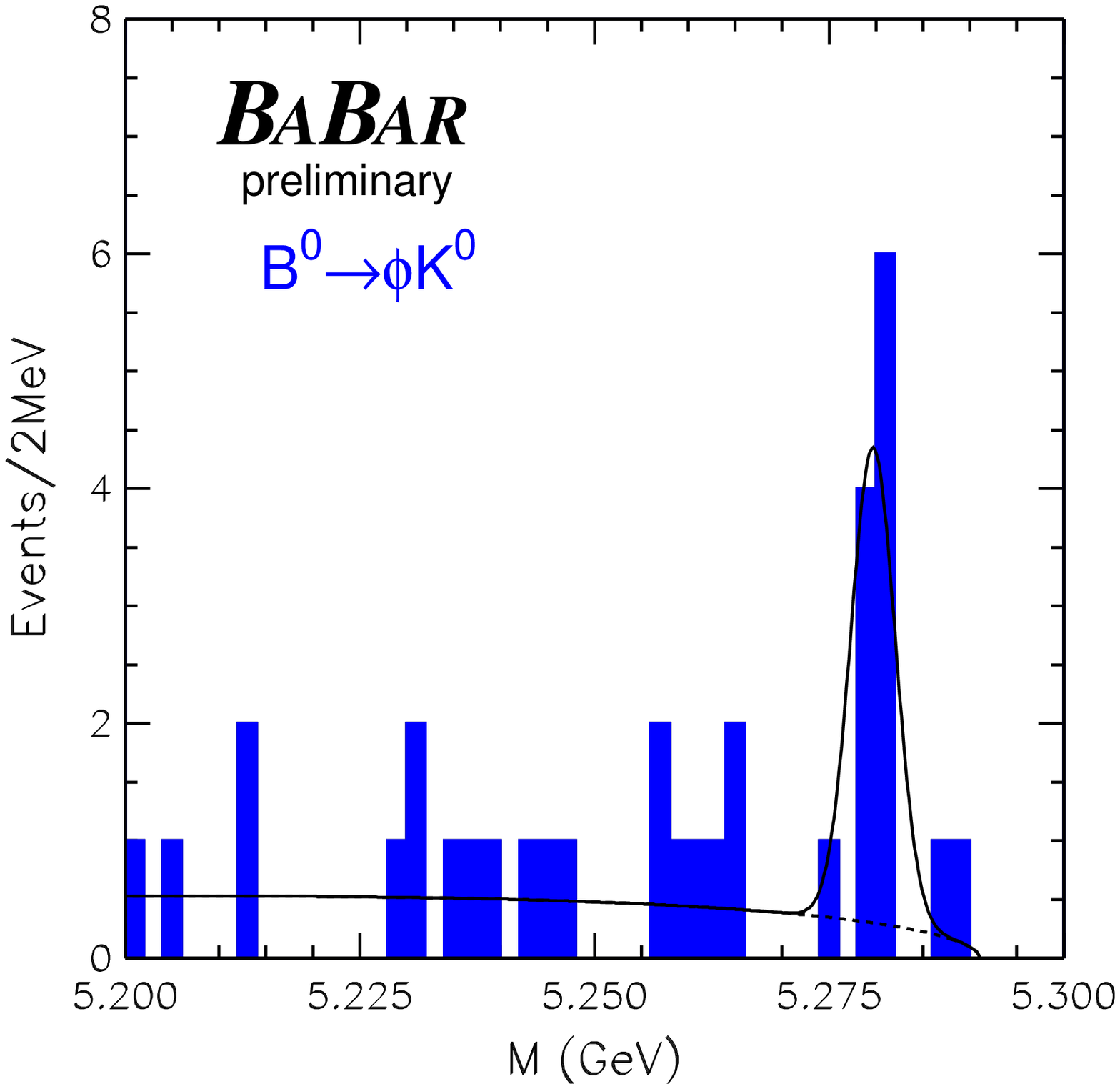}
 \includegraphics[height=0.25\textheight]{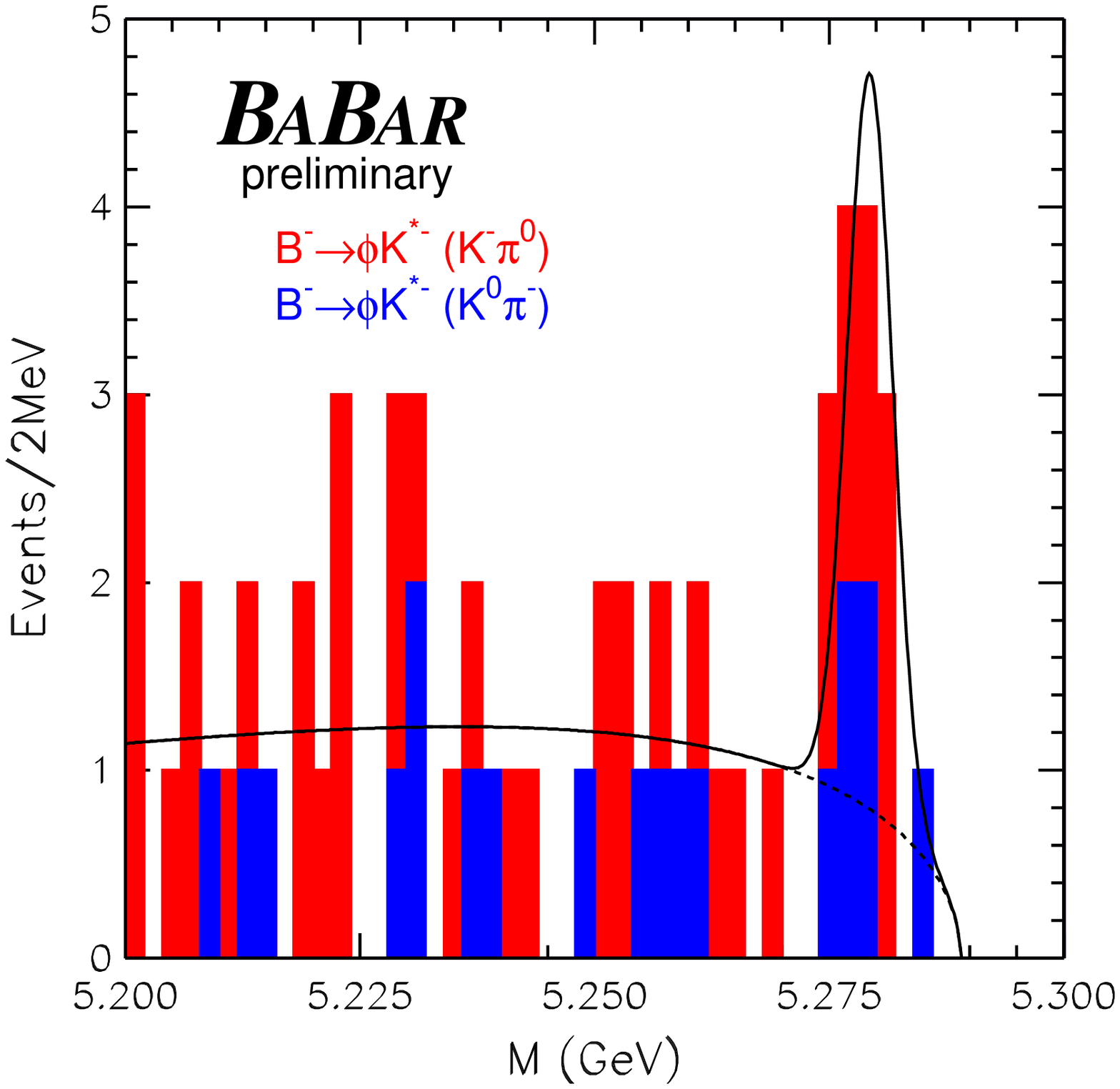}
 \hfill
 \includegraphics[height=0.25\textheight]{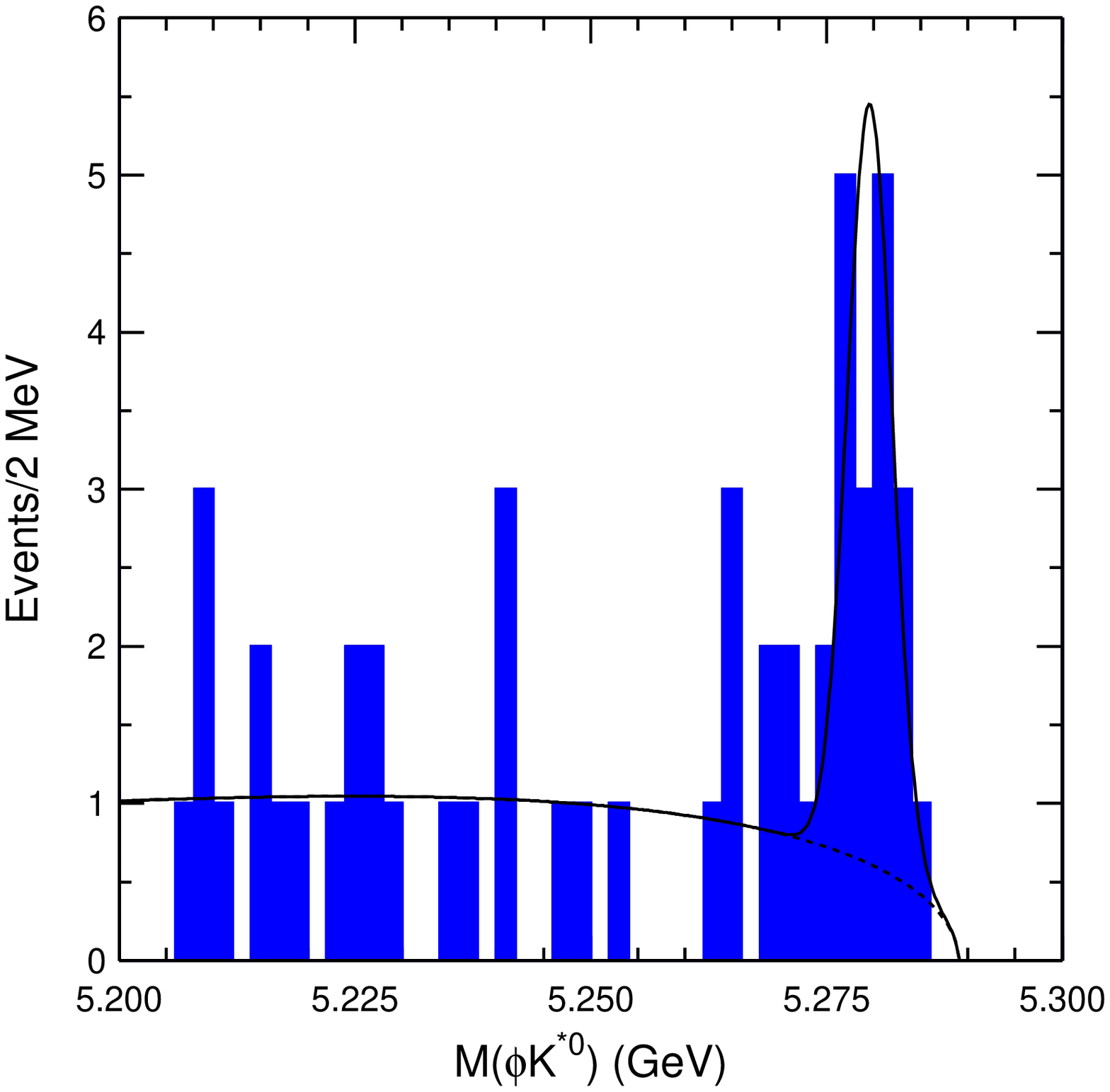}
 \end{center}
  \caption{\it
    The \mes distribution for a) $\Bpm \to \phi \Kpm$, b) $\Bz \to
    \phi \KS$, and c) $\Bpm \to \phi \Kstarpm$, and d) $\Bz \to \phi
    \Kstarz$.  
    Discrete cuts have been applied to enhance the
    signal to background rate for each decay mode.
    \label{fig:phik} }
\end{figure}
time dependent asymmetry in the decay $\bphiks$ will allow an
interesting comparison, especially sensitive to new physics, with the
\stwob measurement in \bpsiks.

\section{Future prospects for \babar}

The first year and half of \babar data has yielded a number of
important results, with the observation of CP violation the most
notable.   The future for \babar looks equally bright; a data sample
of order $500 \invfb$ is expected to be accumulated in the next few
years. This size data sample will enable significant improvements in
accuracy for \stwob, an accurate measurement of $\stwoa_{\rm
  Eff}$, and measurements of the third CP angle $\gamma$.

\section{Acknowledgments}
We are grateful for the extraordinary contributions of our \pep2
colleagues for the superb operation of the B factory. This work is
supported by Department of Energy contract DE-AC03-76SF00515.

\section{References}

\end{document}